# Atomistic Insights into the Degradation of Metal Phthalocyanine Catalysts during Oxygen Reduction Reaction

Huanhuan Yang[1,2,3] and Guangfu Luo[2,3,4,*]

[1]Harbin Institute of Technology, Harbin 150080, P. R. China

[2]State Key Laboratory of Quantum Functional Materials, Department of Materials Science and Engineering, Southern University of Science and Technology, Shenzhen 518055, P. R. China

[3]Guangdong Provincial Key Laboratory of Computational Science and Material Design, Southern University of Science and Technology, Shenzhen 518055, P. R. China

[4]Institute of Innovative Materials, Southern University of Science and Technology, Shenzhen 518055, P. R. China

*E-mail: luogf@sustech.edu.cn

**Abstract**

Oxygen reduction catalysts frequently suffer from degradation under harsh operating conditions, and the limited understanding of the underlying mechanisms hampers the development of effective mitigation strategies. In this study, we integrate first-principles calculations with a time-dependent microkinetic model to investigate the deactivation pathways of six highly active metal phthalocyanines ($M$Pc, $M$ = Cr, Mn, Fe, Ru, Rh, and Ir) during the oxygen reduction reaction (ORR). We quantitatively assess the ORR processes, hydrogen peroxide generation, radical generation, and three primary degradation mechanisms, namely carbon oxidation, nitrogen protonation, and demetallation, through a reaction network involving 40 chemical species and 75 elementary reactions. Our findings reveal that the dominant degradation mechanism varies significantly across the MPcs. Under typical alkaline conditions, the primary byproducts arise from carbon oxidation, driven by •OH radical attack and structural reorganization of surface adsorbates, and from protonation at either the metal center or nitrogen sites. In the kinetics-controlled region, the ORR activity follows the order of RhPc > IrPc > FePc > MnPc > RuPc > CrPc. Notably, RhPc and IrPc demonstrate both higher ORR activity and greater stability than the widely studied FePc under elevated potentials.

**Keywords:** metal phthalocyanines; oxygen reduction reaction; degradation mechanisms; first-principles calculations; microkinetic model

## 1. INTRODUCTION

Metal phthalocyanines (MPcs) represent a prominent class of molecular catalysts characterized by their well-defined $M-N_4$ structure, uniformly distributed active sites, and tunable properties[1]. These features render MPcs both cost-effective and highly efficient for the oxygen reduction reaction (ORR), with potential applications in fuel cells and metal-air batteries[2-5]. Among the examined M-N-C catalysts, Fe-N-C systems have exhibited superior ORR activities compared to Cr-[6], Mn-[7], Co-[7], and Zn-N-C[8] catalysts in both alkaline and acidic conditions[9]. Despite their high catalytic activity, Fe-N-C systems suffer from relatively poor stability in acidic conditions, as evidenced by their larger half-wave potential shifts ($\Delta E_{1/2}$) compared to the Cr[10], Mn[11], Co[12], Zn[8], and Ru[13] counterparts. While Fe-N-C catalysts display improved stability under alkaline conditions[8] (a trend also observed for Mn-N-C[14, 15]), as summarized in **Table S1**, durability still remains a significant challenge. For example, FePc/C shows a significantly higher average current density loss rate (19.2% $h^{-1}$) than Pt/C (13.5% $h^{-1}$) under alkaline conditions[16]. Several strategies have been proposed to enhance the stability, such as substrate modification[17-20], pyrolysis treatment[21], and the use of radial scavenger[22, 23]. However, these efforts still lack robust theoretical guidance due to an incomplete understanding of the underlying deactivation mechanisms.

The degradation of M-N-C catalysts during the ORR has been proposed to occur via three mechanisms. The first is demetallation[24, 25], wherein the central metal ion is replaced by two protons. This process results in the formation of phthalocyanine and the release of metal ions into the solution, a phenomenon experimentally confirmed for FePc under ORR conditions in 0.5 M $H_2SO_4$[26]. The second proposed degradation mechanism involves nitrogen protonation[8, 27, 28]. Although oxidative species may facilitate this mechanism, resulting in M–N bonds cleavage, this process remains thermodynamically unfavorable[28]. The third degradation mechanism is carbon oxidation[12, 29-31], arising either from electrochemical oxidation under high potentials (> 0.8 V *vs.* reversible hydrogen electrode)[32] or from chemical oxidation induced by the highly reactive oxidants during the ORR[33]. For instance, hydrogen peroxide ($H_2O_2$), a 2-electron byproduct during ORR, can be catalyzed by ferrous ions to generate highly reactive hydroxy (•OH) and hydroperoxyl (•OOH) radicals via the Fenton reaction under acidic conditions[22, 34-36]. Previous experiments under $O_2$-saturated 0.1 M $H_2SO_4$ have demonstrated a strong correlation between the content of •OH radicals and the degradation rate of FePc catalyst[33]. Theoretical calculations also showed that the presence of hydroxyl (–OH) or epoxy (C–O–C) groups near the metal sites can reduce the turnover frequency[30]. Nevertheless, the M-N-C ORR catalysts are typically operated under alkaline conditions (pH ~ 13)[17-20], which are short of protons that the above mechanisms rely on. Furthermore, a quantitative assessment of the competition among different degradation mechanisms remains lacking.

Previous theoretical investigations have primarily focused on isolated aspects relevant to the degradation of M-N-C catalysts[37]. For instance, studies have examined the formation energies of M-N-C structures[38], the substitution energies of metal centers by protons[39], as well as the distribution of ORR intermediates and various demetallation states[40] to assess their thermodynamic stability. On the kinetic side, the energy barrier for metal center migration to a neighboring site was calculated[38], while the exceedingly high barriers render such simplified pathways unlikely. Ab initio molecular dynamics (AIMD) simulations have also been employed but no significant structural changes were observed[25, 41]. A common limitation of these studies is their neglect of realistic surface configurations and/or the interplay among multiple degradation pathways under actual operating conditions. In practice, the surface configuration of the catalyst, reaction energy barriers, proton concentration, and the presence of reactive oxygen species vary significantly with pH and potential[42, 43].

In this study, we employ a time-dependent microkinetic modeling, informed by first-principles calculations of 40 species and 75 elementary reactions, to quantitatively evaluate three degradation mechanisms across six superior $M$Pc catalysts ($M$ = Cr, Mn, Fe, Ru, Rh, and Ir) under different pH values and potentials. Our analyses encompass ORR pathways, the generation, conversion, and consumption of reactive oxygen species (ROS), carbon oxidation, nitrogen protonation, and demetallation processes. Our results reveal that under the typical alkaline operating conditions, the formation of major byproducts is primarily from carbon oxidation (driven by •OH radical attack and structural reorganization of surface adsorbates) and from protonation at either the metal center or nitrogen sites, shedding light on the pathways governing MPc catalyst stability. In the kinetics-controlled region, the ORR activity follows the order of RhPc > IrPc > FePc > MnPc > RuPc > CrPc.

## 2. COMPUTATIONAL DETAILS

All first-principles calculations were performed based on density functional theory as implemented in the Gaussian 16 program (Revision C.01)[44], employing with the B3LYP exchange-correlation functional.[45, 46] Van der Waals interactions were incorporated via the Grimme-D3 dispersion scheme.[47] Unrestricted open-shell calculations were utilized to account for potential antiferromagnetic interactions between MPcs and adsorbents. A range of initial geometries and spin multiplicities were carefully examined to identify the most stable configurations. The all-electron 6-311++G** basis sets were used for H, C, N, O, Cr, Mn, and Fe, while the def2TZVP basis sets with effective core potentials were applied to Ru, Rh, and Ir. Solvation effects were included using the polarizable continuum model,[48] assuming water as the solvent with a relative dielectric constant of 78.36. Temperature effects were taken into account by including vibrational and rotational freedoms, and additional translational freedoms for gas species. All calculations were performed at 300 K, while partial pressures of $O_2$ and $H_2$ were set to 1.0 atm. The energy and force tolerances of $2.72 \times 10^{-5}$ eV and 0.023 eV/Å, respectively, and no imaginary frequencies were observed for stable structures. Transition states were located using the Berny algorithm[49] and confirmed by the presence of a single imaginary vibrational frequency associated with the expected reaction coordinate. The pH effects were included to account for proton activity[50, 51] by the equation: $\Delta G_{\text{pH}} = 2.303 \text{k}_\text{B} T \times \text{pH}$.

Since the MPc catalysts are typically deployed under alkaline conditions[21, 52-54], this study focuses on the pH range of 7–14.

Our microkinetic model encompasses a total of 40 chemical species and 75 elementary steps to simulate the ORR processes, $H_2O_2$ generation, radical generation/conversion/consumption, carbon oxidation, nitrogen protonation, and demetallation, as detailed in **Table S2** of the **Supporting Information**. Mass-action kinetics is employed to describe each elementary step[55]. For a reversible reaction, the net reaction rate $r_i$ of elementary step $i$ is expressed as **Eqns. 1** and **2**.

$$r_i = f_i \prod_{j,k} \theta_j^{s_{i,j}} x_k^{s_{i,k}} - b_i \prod_{l,m} \theta_l^{s_{i,l}} x_m^{s_{i,m}} \tag{1}$$

$$\frac{\partial \theta_j}{\partial t} = \sum_i s_{i,j} r_i \ , \tag{2}$$

where $f_i$ and $b_i$ are the forward and backward rate constants, respectively; $\theta_j$ is the surface coverage of adsorbed species $j$; $x_k$ is the dimensionless concentration of ROS $k$ normalized to 1 mol/L; and $s_{i,j}$ is stoichiometry coefficient of species $j$ in elementary step $i$. The use of dimensionless concentrations for ROS simplifies the physical model and has been adopted in previous studies[56, 57]. Given the inherently non-equilibrium nature of degradation process, a time-dependent simulation is performed using an in-house Mathematica code, which is provided in the **Supporting Information**. Further details on the time-dependent differential equations, the forward and backward rate constants, and the Gibbs free energy changes and forward barriers are provided in the **Supporting Information**.

## 3. RESULTS AND DISCUSSION

To quantitatively evaluate the degradation mechanisms of six MPc catalysts ($M$ = Cr, Mn, Fe, Ru, Rh, and Ir), which were previously predicted with high catalytic activity for ORR[58], we sequentially elaborate the ORR processes and degradation mechanisms, reaction network, time evolution of reaction species under different pH values and potentials, and reaction network and dominant degradation products under typical operating conditions.

### 3.1 ORR processes and degradation mechanisms

It is well-established that the ORR processes can proceed via a 4-electron (4e⁻) pathway involving four intermediates ($^MO_2$, $^MOOH$, $^MO$, and $^MOH$), as well as a 2-electron (2e⁻) pathway involving three intermediates ($^MO_2$, $^MOOH$, and $^MH_2O_2$)[59] that lead to a partially reduced product, $H_2O_2$, as illustrated in **Figure 1a**. The corresponding free energy diagrams are provided in **Figure S1**. According to our microkinetic model with only the two pathways, the primary ORR-related species are the bare catalyst (denoted as $^M$), $^MOH$, and $^MH_2O_2$, as exemplified by FePc in **Figure 1b**. Results for all examined MPcs and computational details are provided in **Figure S2** and **Section 4** of **Supporting Information**, respectively.

The generation and conversion of ROS occur through multiple reaction pathways (**Figure 1a**).

Specifically, the •OH radical can be produced via dissociation of the O–O bond in $^M$OOH and $^M$H$_2$O$_2$, or through the decomposition of aqueous H$_2$O$_2$[60, 61]. The •OH radical can further react with aqueous H$_2$O$_2$ to yield the •OOH radical[62, 63], namely •OH + H$_2$O$_2$ (*aq*) → •OOH + H$_2$O (*l*). Conversely, •OOH can react with aqueous H$_2$O$_2$ to regenerate •OH: •OOH + H$_2$O$_2$ (*aq*) → •OH + H$_2$O (*l*) + O$_2$ (*g*). Moreover, the adsorbed hydrogen (*H) is known to activate H$_2$O$_2$, cleaving the peroxide bridge to generate •OH[64]: *H + H$_2$O$_2$ (*aq*) → * + •OH + H$_2$O (*l*). A total of 29 reactions related to ROS generation and conversion are detailed in **Table S4**.

We investigate three primary degradation mechanisms of MPc catalysts during ORR: carbon oxidation, nitrogen protonation, and demetallation, as illustrated in **Figure 1c**. Carbon oxidization involves the attack of •OH or •OOH radicals on carbon atoms near the metal center, leading to surface oxidation or embedding of oxygen into the carbon matrix. Nitrogen protonation refers to the addition of protons to the nitrogen sites, as potentially destabilizes the coordination of M–N. Since direct metal-leaching is highly unfavorable (**Table S5**), we propose an alternative pathway. This mechanism initiates with oxidation of the metal center M by •OH radicals, which destabilize the metal center, followed by nitrogen protonation to stabilize the demetallated structure, and ultimately, release of the M(OH)$_x$ into solution.

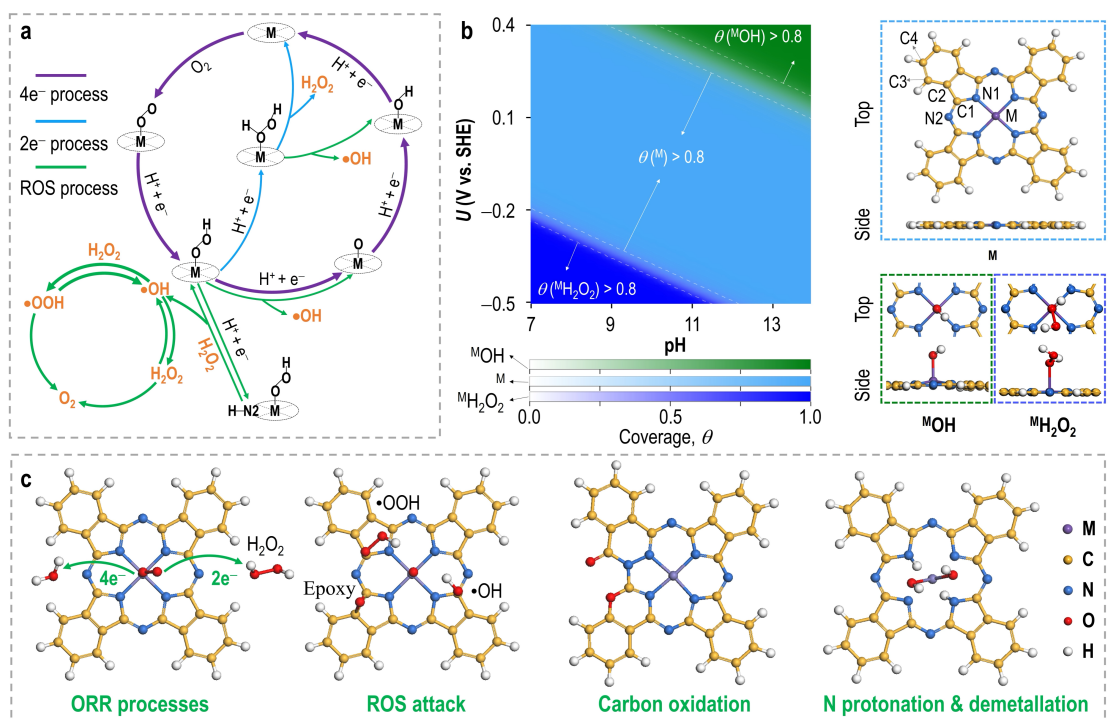

**Figure 1.** (a) ORR pathways and the generation/conversion of ROS; (b) Pourbaix map illustrating the coverage of primary ORR intermediates on FePc, along with the site labelling scheme and top/side views of three primary states. Only the 2e$^-$ and 4e$^-$ ORR pathways are considered here. (c) Illustration of ORR processes, ROS attack, and three degradation mechanisms: carbon oxidation, nitrogen protonation, and demetallation.

## 3.2 Reaction network

We construct a reaction network for MPcs that integrate the ORR pathways, the interactions between the ORR intermediates and two radicals •OH and •OOH, as well as three degradation mechanisms, as illustrated in **Figure 2**. In total, the reaction network involves 40 chemical species and 75 elementary reactions. These species are classified into four categories: (1) ORR-related intermediates, including $^{M}O_2$, $^{M}OOH$, $^{M}O$, $^{M}OH$, and $^{M}H_2O_2$, where $^{M}$ stands for adsorption site at metal center; (2) $O_2$ and ROS ($H_2O_2$, •OH, and •OOH), which exist in solution and participate in multiple reactions; (3) Byproducts-I, consisting of 25 states characterized by adsorbates (*$O_2$, *OOH, *O, *OH, *$H_2O_2$, and *H) bound to M, C, and N sites; (4) Byproducts-II, comprising $O_{inter}$, $^{M}OH-O_{inter}$, $^{M}O-O_{outer}$, Pc+M(OH)$_2$, and Pc+M(OH)$_3$, which involve significant structural transformations, namely, C–C/C–N bond cleavage, C–O/N–N bond formation, or the metal center loss.

This network comprises four types of elementary reactions: (1) proton-coupled electron transfer reactions, (2) •OH and •OOH attacks on metal and carbon sites, (3) adsorption/desorption of $O_2$, water, and peroxide, and (4) dissociation of O–O bond or structural reorganization of intermediates, the last of which are inspired by our molecular dynamics simulations. Reactions occurring entirely in solution, without direct involvement of MPcs, are not shown in **Figure 2** but are presented in **Figure 1a** and **Table S2**.

In this network, the ORR-related intermediates are represented by circular nodes, and the 4e⁻ pathway are arranged in a pentagonal network $1 \to 2 \to 3 \to 4 \to 5 \to 1$, while the 2e⁻ pathway proceeds through the network $1 \to 2 \to 3 \to 6 \to 1$.

Regarding carbon oxidation, our calculations reveal that the carbon atoms near the metal center (C1 site, **Figure 1b**) are the most susceptible to oxidation, while the C2 site exhibit the highest stability (see relative energies in **Table S6**). Consequently, we investigate states related to the radical attack of the C1 site, namely, $^{C1}OH$ and $^{C1}OOH$. For $^{C1}OH$-related species, we examine five surface states: $^{C1}OH$ (state 11), $^{M}O_2-^{C1}OH$ (state 7), $^{M}OOH-^{C1}OH$ (state 8), $^{M}O-^{C1}OH$ (state 9), and $^{M}OH-^{C1}OH$ (state 10). These states interconvert through the pathway $11 \to 7 \to 8 \to 9 \to 10 \to 11$. Additionally, they can be formed from or changed into the ORR-related intermediates via radical attack or hydrogenation reactions. States 8, 9, and 10 can undergo hydrogen transfer between the sites M and C1, forming $^{M}H_2O_2-^{C1}O$ (state 21), $^{M}OH-^{C1}O$ (state 15), and $^{M}H_2O-^{C1}O$ (state 20), respectively. When the $^{C1}O$ rotates outward ~90°, it adopts the $O_{outer}$ configuration (state 19). If it rotates ~90° and inserts into the C1–C2 bond of the five-membered pyrrole ring (C$_4$N), it forms two $O_{inter}$ configurations (states 13 and 14). Because the interaction between •OOH radical and the C1 site to form $^{C1}OOH$ is approximately 1.1 eV less favorable than that of •OH radical forming $^{C1}OH$ (**Table S7**), a trend similar to the relative stability between $^{M}OOH$ and $^{M}OH$, we examine only three major states involving $^{C1}OOH$: $^{M}O-^{C1}OOH$ (state 16), $^{M}OH-^{C1}OOH$ (state 17), and $^{M}O-^{C1}O$ (state 18), given the higher energy level of $^{C1}OOH$, $^{M}O_2-^{C1}OOH$, and $^{M}OOH-^{C1}OOH$. They can be formed and undergo transformation via the pathway 4 → 16 (5) → 17 → 18 → 19.

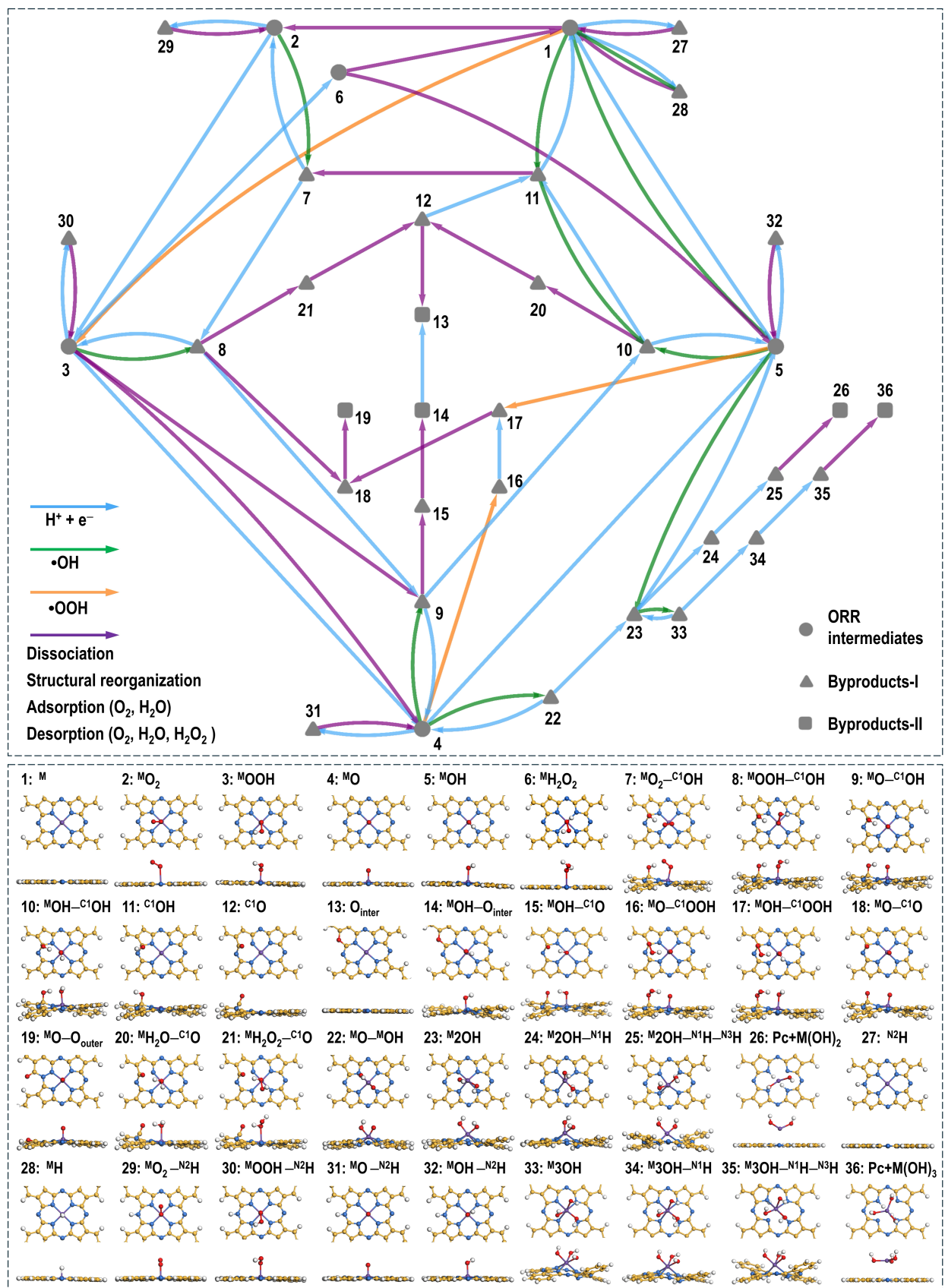

**Figure 2.** Reaction network of MPcs during ORR, with inclusion of the carbon oxidation, nitrogen protonation, and demetallation. The upper panel depicts the reaction network, with arrows color-coded by reaction type and node shapes indicating species categories. The lower panel presents the corresponding structures, with top and side views aligned vertically.

With respect to protonation, the protonation at site N1 may break the M–N bonds during the ORR, potentially leading to demetallation through the reaction[58]: MPc ($aq$) + $n$H$^+$($aq$) → M$^{n+}$($aq$) + Pc ($aq$) + ($n$-2)/2 H$_2$ ($g$). However, our calculations indicate that the protonation at site N1 is thermodynamically unfavorable for bare MPc (**Table S8**), requiring an energy input of over 1.0 eV at 0 V *vs.* SHE, consistent with previous reports[25, 28]. In contrast, protonation at site N2 is more favorable (state 27), requiring an energy input of no more than 0.5 eV. Particularly, this process is exothermic for RuPc, RhPc, and IrPc. The nitrogen protonation can also occur on the ORR-related intermediates, forming $^M$O$_2$–$^{N2}$H (state 29), $^M$OOH–$^{N2}$H (state 30), $^M$O–$^{N2}$H (state 31), and $^M$OH–$^{N2}$H (state 32). Additionally, we incorporate the protonation at the metal site into the network (state 28), since this process is exothermic for RhPc and IrPc.

The demetallation mechanism can initiate from •OH radical attacks on three primary states $^M$, $^M$OH, and $^M$O, forming $^M$OH, $^M$2OH (state 23), and $^M$O–$^M$OH (state 22), respectively. State $^M$O–$^M$OH can undergo further hydrogenation to yield $^M$O (state 4) or $^M$2OH (state 23). If the N sites adjacent to the metal center are subsequently protonated, the metal ion can be released, following the reaction pathway 23 → 24 → 25 → 26. Additionally, •OH attack on state 23 may produce another byproduct, $^M$3OH (state 33), due to the stability of high-valence states of certain metal centers. Consequently, states 24, 25, and 26 are replaced by $^M$3OH–$^{N1}$H (state 34), $^M$3OH–$^{N1}$H–$^{N3}$H (state 35), and Pc+M(OH)$_3$ (state 36). For state 22 to 25 and 33 to 35, the metal center protrudes from the Pc plane due to the significant structural distortion.

**3.3 Time evolution of reaction species under different pH values and potentials**

Using the constructed reaction network, we calculate the thermodynamic and kinetic properties of the elementary steps (see details in **Supplementary Materials**). These results are integrated into a time-dependent microkinetic model that accounts for variation in pH and applied potential (see **Table S2** for details**)**. The initial condition assumes a clean catalyst surface, free of adsorbed species. Since ORR catalysts usually operate under neutral to alkaline conditions, particularly at pH 13 and within a potential range of 0.2 to 1.1 V *vs.* RHE (0.5 to 0.8 V *vs.* RHE is required to reach the limiting current density[11]), we investigate how the primary surface species evolve over time as a function of pH and potential.

**Figure 3a** illustrates the time evolution of dominant species on FePc across a pH range of 7 to 14 at a potential of −0.1 V *vs.* SHE, corresponding to 0.22 to 0.73 V *vs.* RHE based on the equation[65]: $U_{RHE}$ = $U_{SHE}$ + 2.303k$_B$$T$ × pH / e. Under these conditions, FePc predominantly exists in its bare surface state.



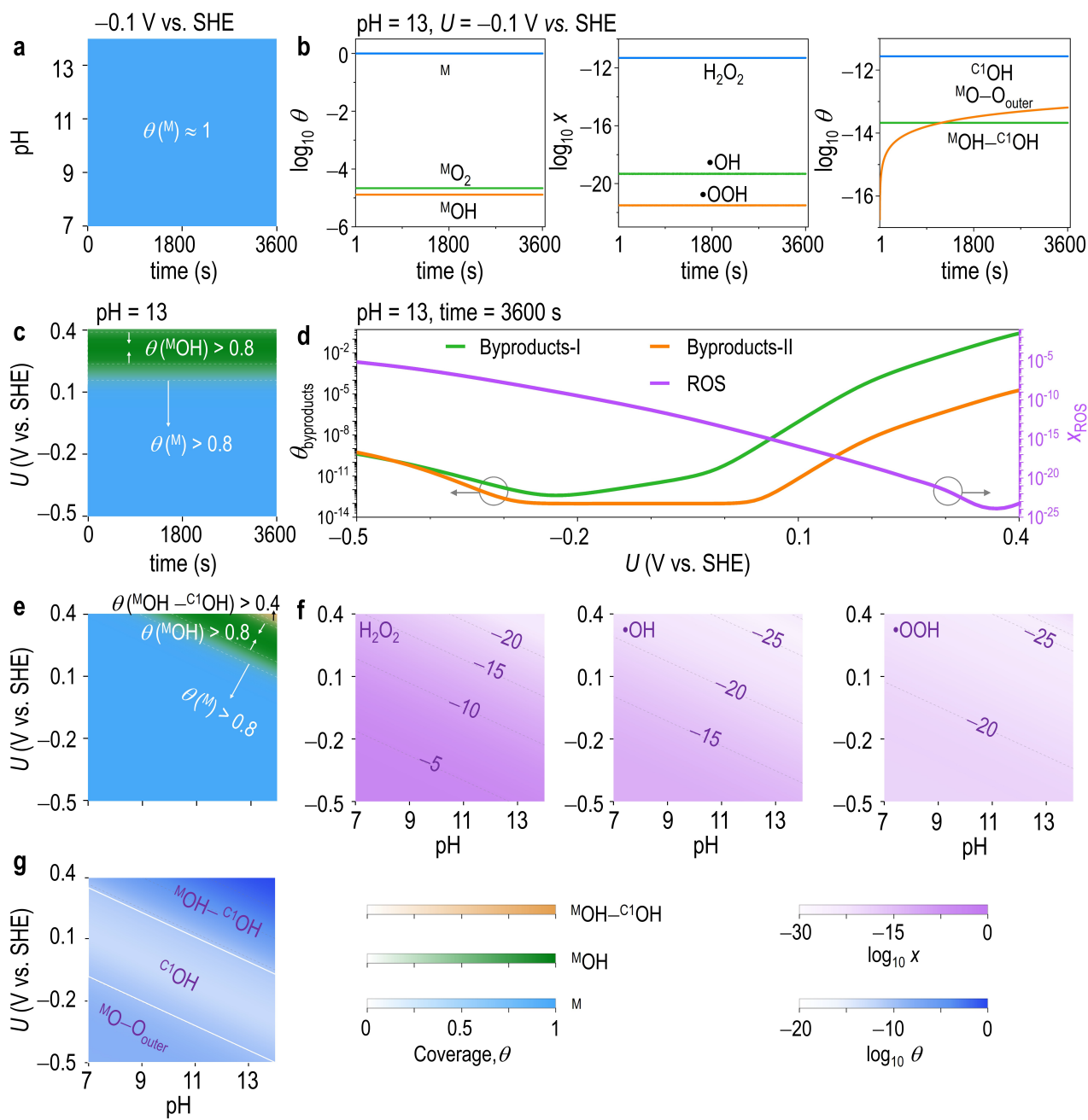

**Figure 3.** Primary ORR intermediates, ROS, and byproducts on FePc. (a) Time evolution of primary ORR intermediates as a function of pH at −0.1 V vs. SHE. (b) Time evolution of ORR intermediates, ROS, and byproducts at pH 13 and −0.1 V vs. SHE (0.67 V vs. RHE). (c) Time evolution of primary ORR intermediates as a function of potential at pH 13. (d) Total concentrations of ROS, and byproducts as a function of potential at pH 13 after 3600 s. (e-g) Pourbaix diagrams illustrating the distribution of (e) primary surface states, (f) ROS, and (g) major byproducts as a function of pH and potential after 3600 s. See results of individual byproducts in **Figure S3a** and **Figure S4a**. The concentrations of ORR intermediates and byproducts are in the unit of number per MPc molecule, while $x$ denotes the dimensionless concentration normalized to 1 mol/L.



**Figure 3b** further presents the time evolution of other major ORR-related intermediates, ROS, and byproducts at pH 13 and −0.1 V *vs.* SHE (0.67 V *vs.* RHE), a typical ORR operating condition. We find that all major ORR-related intermediates stabilize in less than $10^{-5}$ seconds, with $^{M}O_2$ identified as the second most abundant surface species, followed by $^{M}OH$. Among the ROS, $H_2O_2$ is the dominant species, while the concentrations of •OH and •OOH radicals are several orders of magnitude lower. Notably, the •OH radical primarily arises from the dissociation of $^{M}OOH$ to $^{M}O$ and $^{M}H_2O_2$ to $^{M}OH$, with only a minor contribution from the reaction between *H and $H_2O_2$, as previously suggested[64]. This behavior also differs from that in acidic environments, where the Fenton reaction drives the conversion of $H_2O_2$ into •OH and •OOH radicals[22] but becomes largely inactivate under alkaline conditions. The most prominent byproduct is $^{C1}OH$, followed by $^{M}O–O_{outer}$, and $^{M}OH–^{C1}OH$.

Our results reveal that the applied potential plays a critical role in determining the dominant surface states, ROS levels, and byproduct formation. As shown in **Figure 3c**, the dominant ORR species transitions from $^{M}$ to $^{M}OH$ as the potential increases above 0.2 V *vs.* SHE at pH 13. Within the potential range of −0.5 to 0.4 V *vs.* SHE, the total ROS concentration decreases significantly with potential, from approximately $10^{-6}$ to $10^{-24}$ (**Figure 3d**). Concurrently, the total concentrations of byproducts exhibit a "V"-shaped trend, reaching a minimum near −0.2 V *vs.* SHE.

We further present a Pourbaix diagram to elucidate the primary ORR species, ROS, and byproducts for FePc over the pH range of 7−14 and potentials from −0.5 to 0.4 V *vs.* SHE after 3600 s. As shown in **Figure 3e**, $^{M}$ and $^{M}OH$ are the primary ORR intermediates under most conditions. However, at high pH and elevated potentials (pH > 13 and $U$ > 0.3 V *vs.* SHE), the state $^{M}OH–^{C1}OH$ emerges as the predominant surface species. Compared to the simulation without degradation mechanisms (**Figure 1b**), the concentration of $^{M}H_2O_2$ is significantly reduced in the region of $U < 0.21 - 0.06\text{pH}$, primarily due to the dissociation of $^{M}H_2O_2$ to $^{M}OH$ (6 → 5), or desorption to $^{M}$ (6 → 1), as we will elaborate later.

The dominant ROS is $H_2O_2$, the concentration of which is 5 to 10 orders higher than those of •OH and •OOH (**Figure 3f**). The ROS concentrations peak under conditions of low potential and low pH. Under the examined conditions, $^{M}OH–^{C1}OH$, $^{C1}OH$, and $^{M}O–O_{outer}$ are identified as the primary byproducts for FePc (**Figure 3g**). Among them, $^{M}OH–^{C1}OH$ is the most abundant species when $0.78 - 0.06\text{pH} < U$, indicating substantial electrochemical oxidation at the C1 site. In the intermediate range of $0.34 - 0.06\text{pH} < U < 0.78 - 0.06\text{pH}$, $^{C1}OH$ becomes the dominant byproduct, whereas byproducts-II $^{M}O–O_{outer}$ predominates under the remaining conditions. These findings clearly indicate that carbon oxidation dominates the degradation pathway for FePc, whereas demetallation remains negligible under the investigated conditions.



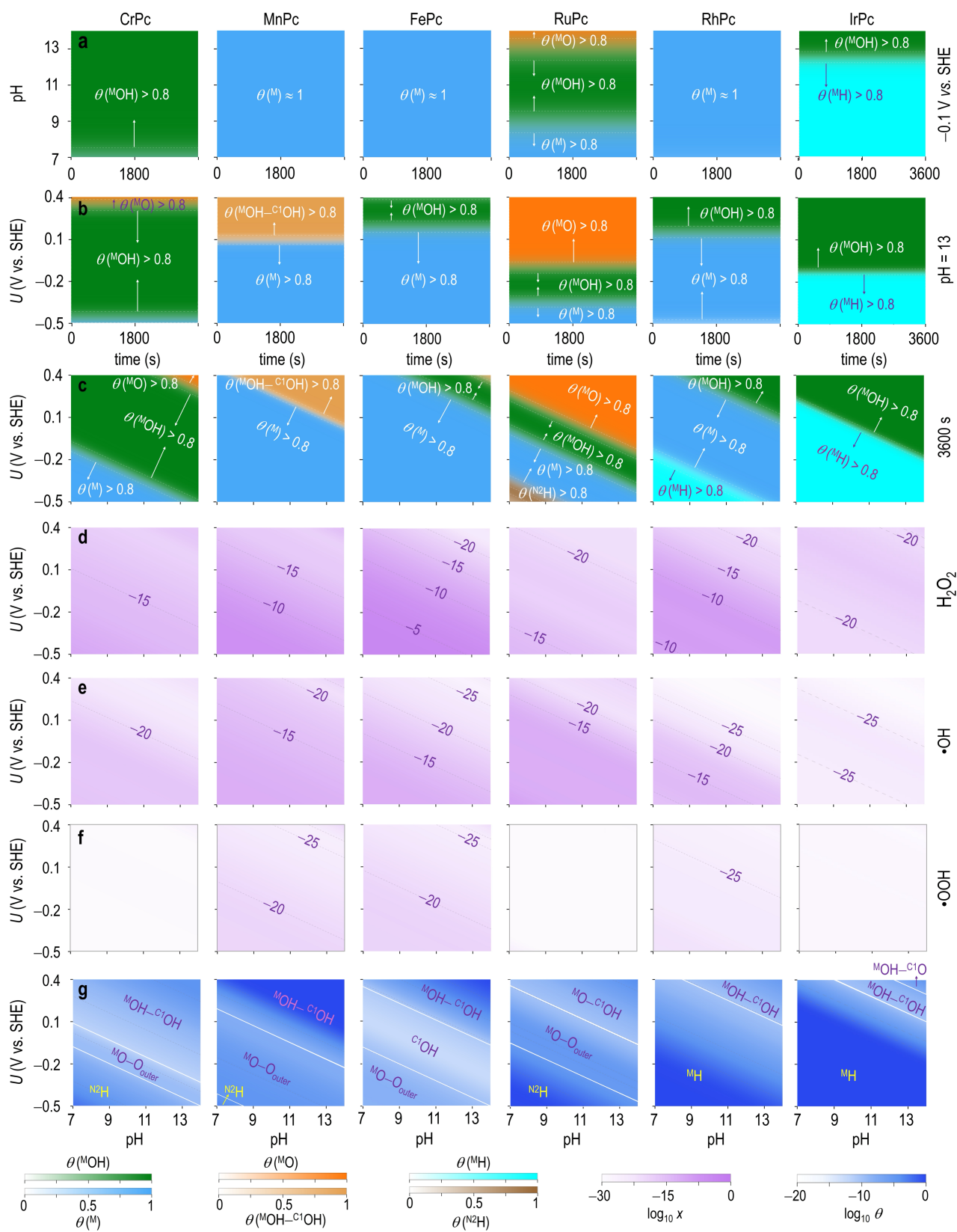

**Figure 4.** Primary ORR intermediates, ROS, and major byproducts for six MPcs ($M$ = Cr, Mn, Fe, Ru, Rh, and Ir). (a) Time evolution of dominant surface states as a function of pH at −0.1 V *vs.* SHE; (b) Time evolution of dominant surface states as a function of potential at pH 13; Pourbaix diagrams for the (c) dominate surface states, (d-f) ROS, and (g) dominant byproducts after 3600 s. See results of individual byproducts in **Figures S3-S4**.



Building on the insights gained from FePc, we first conduct time-resolved analyses for five other MPcs: CrPc, MnPc, RuPc, RhPc, and IrPc. At a representative potential of −0.1 V *vs.* SHE, the dominant species are $^M$OH for CrPc; $^M$ for MnPc; $^M$, $^M$OH, and $^M$O for RuPc; $^M$ for RhPc; and $^M$H and $^M$OH for IrPc, across the pH range of 7–14 (**Figure 4a**). Notably, byproduct $^M$H appears as a dominant species for IrPc at pH below 12. Under the typical alkaline condition of pH 13 (**Figure 4b**), CrPc, FePc, RuPc, and RhPc are predominantly covered by ORR-related intermediates within the whole examined potential range. In contrast, MnPc begins to accumulate the byproduct $^M$OH–$^{C1}$OH at $U > 0.1$ V, while IrPc forms $^M$H at $U <$ −0.1 V, respectively.

The Pourbaix diagrams in **Figure 4c** reveal that $^M$O, $^M$OH, and $^M$ are three major ORR species among the MPcs. Surprisingly, byproducts $^M$OH–$^{C1}$OH, $^{N2}$H, and $^M$H emerge as dominant species under specific conditions. Specifically, byproduct $^M$OH–$^{C1}$OH dominates MnPc in the region of $U > 0.89 − 0.06$pH; $^{N2}$H becomes prevalent for RuPc when $U < 0.1 − 0.06$pH; and $^M$H prevails for RhPc and IrPc under conditions of $U < 0.28 − 0.06$pH and $U < 0.64 − 0.06$pH, respectively. These trends contrast sharply to the results obtained without considering the degradation processes (**Figure S2**).

Regarding ROS generation, $H_2O_2$ and •OH are the major ROS across all systems (**Figures 4d-f**), while MnPc and FePc uniquely produce higher amounts of •OOH compared to the other four MPcs. Under typical experimental conditions of −0.1 V *vs.* SHE and pH 13, the total ROS concentration follows the order of FePc > RhPc > MnPc > RuPc > CrPc > IrPc.

Finally, we examine the dominant byproducts of these MPcs under various operating conditions (**Figure 4g**). In general, byproducts associated with carbon oxidation, including $^M$OH–$^{C1}$OH, $^M$O–O$_{outer}$, $^M$O–$^{C1}$OH, and $^M$OH–$^{C1}$O, tend to form under relatively high potentials and high pH, whereas protonation byproducts ($^M$H and $^{N2}$H) are more prevalent under low potentials and low pH conditions. Specifically, the dominant byproducts are $^{N2}$H, $^M$O–O$_{outer}$, and $^M$OH–$^{C1}$OH for CrPc and MnPc; $^{N2}$H, $^M$O–O$_{outer}$, and $^M$O–$^{C1}$OH for RuPc; $^M$H and $^M$OH–$^{C1}$OH for RhPc; and $^M$H, $^M$OH–$^{C1}$OH, and $^M$OH–$^{C1}$O for IrPc. Under the typical operating conditions of −0.1 V *vs.* SHE and pH 13, the dominant byproducts are $^M$OH–$^{C1}$OH for CrPc and MnPc, $^{C1}$OH for FePc, $^M$O–O$_{outer}$ for RuPc, $^M$H for RhPc and IrPc.

### 3.4 Reaction network under typical operating conditions

To intuitively elucidate the degradation processes of MPcs under typical operating conditions (pH 13 and −0.1 V *vs.* SHE after 3600 s), we adjust the line thickness and node size in the reaction network according to the net reaction rates and species concentrations, respectively (**Figure 5**). To emphasize the key processes and species, reactions with negligible net rates and species with very low concentrations are omitted. Notably, certain hydrogenation steps are found to exhibit reverse net reaction direction, majorly because of the extremely low concentrations of their reactants, and are highlighted with red lines.



It's evident that the 4e⁻ ORR pathway (1 → 2 → 3 → 4 → 5 → 1) is substantially more favorable than the 2e⁻ pathway (1 → 2 → 3 → 6 → 1). Since the step 1 → 2 corresponds to $O_2$ consumption, it serves as an indicator of overall catalytic activity. Among the six MPcs, the $O_2$ consumption rates follow the trend of IrPc (5.2 s$^{-1}$) > RhPc (3.8 s$^{-1}$) > RuPc (9.6 × 10$^{-2}$ s$^{-1}$) > MnPc ≈ FePc (3.5 × 10$^{-2}$ s$^{-1}$) > CrPc (1.8 × 10$^{-3}$ s$^{-1}$).

Among all the MPcs, CrPc exhibits the simplest network, primarily dominated by the 4e⁻ ORR pathway (**Figure 5a**; full network shown in **Figure S5**). Three other major processes include dissociation via $^M$OOH → $^M$O + •OH (3 → 4), •OH radical attack through $^M$OH + •OH → $^M$OH–$^{Cl}$OH (5 → 10), and protonation of state 10 to regenerate state 5: $^M$OH–$^{Cl}$OH + H$^+$ + e$^-$ → $^M$OH + $H_2O$. The primary ORR intermediate is $^M$OH, while $^M$OH–$^{Cl}$OH is the main byproduct.

The reaction network of IrPc (**Figure 5b**; full network shown in **Figure S6**) is dominated by the 4e⁻ ORR process and a secondary route involving $^{Cl}$OH and $^M$OH–$^{Cl}$OH states: $^M$OH → $^M$OH–$^{Cl}$OH → $^{Cl}$OH → $^M$ (5 → 10 → 11 → 1). The primary ORR intermediates in this system include $^M$OH, $^M$, and $^M$O$_2$. The main byproduct is $^M$H (state 28), which even dominates the surface state of IrPc and blocks ORR reaction under low potentials, as we will show in subsequent context.

For RhPc (**Figure 5c**; full network shown in **Figure S7**), the network is still dominated by the 4e⁻ ORR pathway, accompanying a secondary pathway related to $O_2$ consumption on $^{Cl}$OH surface: $^M$ → $^{Cl}$OH → $^M$O$_2$–$^{Cl}$OH → $^M$O$_2$ (1 → 11 → 7 → 2). The primary ORR intermediates in this system are $^M$, $^M$OH, and $^M$O$_2$. $^M$H is identified as the main byproduct, similar to IrPc.

As the most extensively studied system, FePc exhibits a reaction network that combines partial features observed in CrPc, IrPc, and RhPc (**Figure 5d**; full network shown in **Figure S8**). In addition, a new pathway, $^M$OOH → $^M$H$_2$O$_2$ → $^M$OH (3 → 6 → 5) emerges, contributing to •OH radical formation. This pathway leads to a reduced $^M$H$_2$O$_2$ concentration compared to the simulation without degradation mechanisms (**Figure 1b**). Furthermore, •OH radical attacks appear in two reactions: $^M$ + •OH → $^{Cl}$OH (1 → 11) and $^M$ + •OH → $^M$OH (1 → 5), although their contributions remain relatively minor. Another relatively low-rate pathway involves $^M$O–$^{Cl}$OH and $^M$OH–$^{Cl}$OH states, proceeding through 3 → 9 → 10 → 5. Similar to IrPc and RhPc, the dominant ORR intermediates are $^M$, $^M$O$_2$, and $^M$OH. The primary byproduct is $^{Cl}$OH, while its concentration is significantly lower than that in other MPcs.



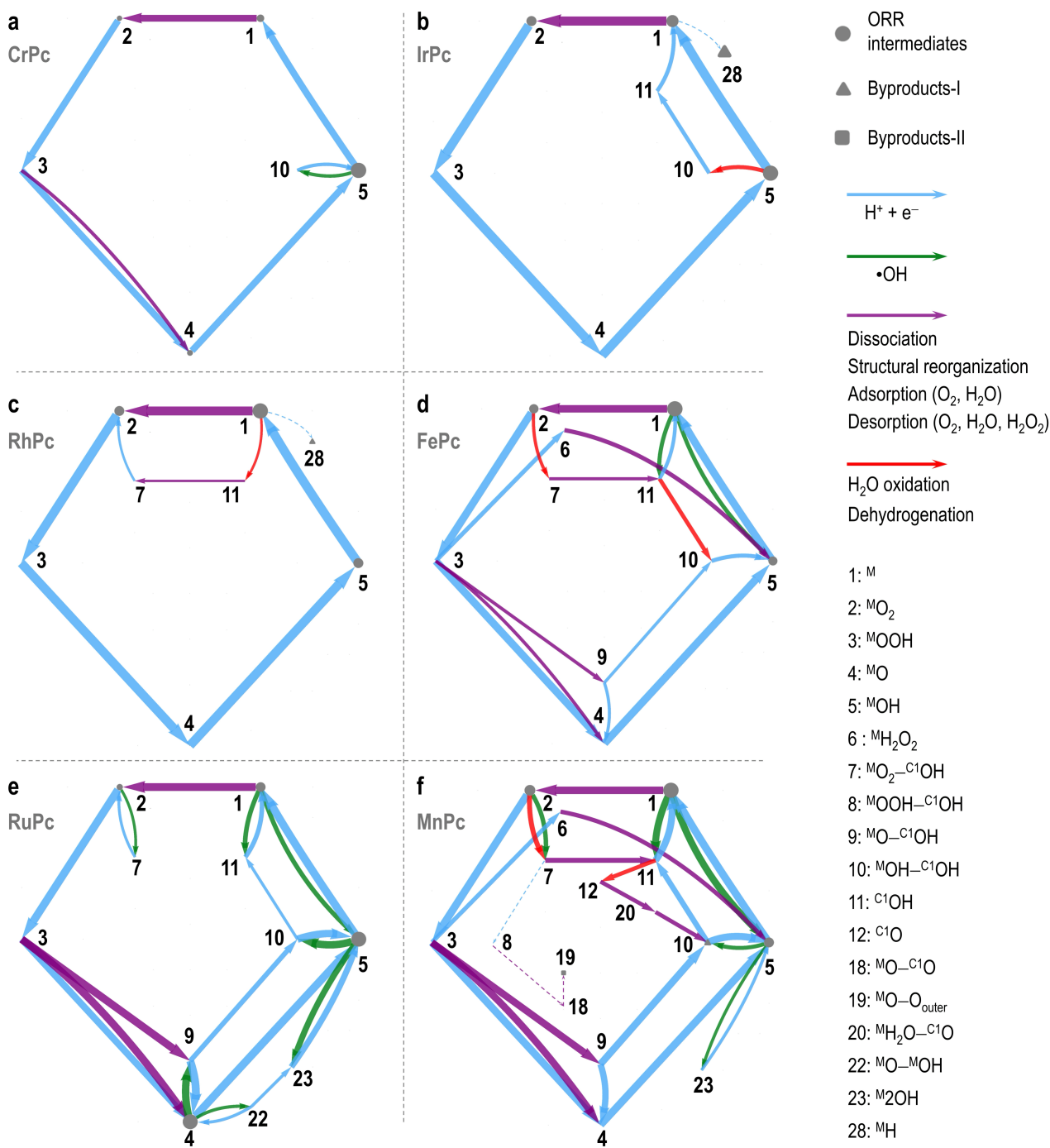

**Figure 5.** Reaction networks of MPcs during ORR at pH 13 and −0.1 V *vs*. SHE after 3600 s. The networks are arranged in order of increasing complexity: (a) CrPc, (b) IrPc, (c) RhPc, (d) FePc, (e) RuPc, and (f) MnPc. Line thickness and node size represent the net reaction rate $r$ and surface coverage $\theta$, scaled by 2 × $\log_{10}|r|$ + 25 and 0.06 × $\log_{10}\theta$ + 0.7, respectively. For visualization clarity, $|r|$ and $\theta$ below $10^{-10}$ are assigned fixed minimum values: a line thickness of 2 and a fixed size of 0.1. To highlight key features, reactions with $|r|$ below $5 \times 10^{-10}$ s$^{-1}$ and species with $\theta$ below $10^{-8}$ per MPc are omitted. However, for reactions with $|r|$ below $5 \times 10^{-10}$ s$^{-1}$ that led to species with $\theta$ above $10^{-8}$ per MPc, dashed lines are added to indicate the corresponding reaction pathways.



In the case of RuPc (**Figure 5e**; full network shown in **Figure S9**), the conventional 4e⁻ ORR process remains a primary component of the reaction network, with $^M$O and $^M$OH as the dominant ORR intermediates. However, two dissociation pathways exhibit reaction rates as prominent as the 4e⁻ ORR pathway: one occurs through $^M$OOH → $^M$O–$^{C1}$OH (3 → 9), and the other via $^M$OOH → $^M$O + •OH (3 → 4). This leads to a relatively high concentration of •OH, as observed in **Figure 4e**. Additionally, •OH radical attacks contribute significantly to a range of processes, including $^M$ + •OH → $^M$OH or $^{C1}$OH (1 → 5, 11), $^M$OH + •OH → $^M$OH–$^{C1}$OH or $^M$2OH (5 → 10, 23), and $^M$O + •OH → $^M$O–$^{C1}$OH (4 → 9). These results highlight the significant role of •OH radical attack in this system. Although $^M$2OH (state 23) is formed at a high rate, the protonation process significantly consumes the byproducts (23 → 5).

Among the six MPcs, MnPc displays the most complex reaction network, combing the major pathways observed in both FePc and RuPc (**Figure 5f**; full network shown in **Figure S10**). The high concentration of •OH radical (**Figure 4e**) is primarily generated through the reaction $^M$OOH → $^M$O + •OH (3 → 4), followed by $^M$H$_2$O$_2$ → $^M$OH + •OH (6 → 5). These radicals are subsequently consumed via the reactions $^M$ + •OH → $^{C1}$OH (1 → 11) and $^M$ + •OH → $^M$OH (1 → 5), both of which are considerably more prominent than in the FePc and RuPc. Moreover, the byproduct $^M$O–O$_{outer}$ (state 19) accumulates through a relatively low-rate pathway (2 → 7 → 8 → 18 → 19), which initiates with •OH attack on $^M$O$_2$ to form $^M$O$_2$–$^{C1}$OH (state 7), followed by hydrogenation to $^M$OOH–$^{C1}$OH (state 8), dehydration to $^M$O–$^{C1}$O (state 18), and final conversion to $^M$O–O$_{outer}$ via C–N bond cleavage and N–N bond formation. The dominant ORR-related states are $^M$, $^M$OH, and $^M$O$_2$, while $^M$OH–$^{C1}$OH and $^M$O–O$_{outer}$ are identified as the main byproducts.

In addition to the detailed analyses at −0.1 V *vs.* SHE, we finally examine the potential-dependent byproduct content and oxygen consumption rate for all MPcs at pH 13 after 3600 s. The two metrics can be considered indicators of degradation rate and ORR activity, respectively. As illustrated in **Figure 6a**, the total byproduct contents generally follow a "V"-shaped trend, indicating the existence of an optimal voltage for the stability of each MPc. Specifically, the minimum byproduct content is observed at −0.24 V for FePc, −0.20 V for CrPc, 0.00 V for RuPc, 0.13 V for RhPc, 0.18 V for IrPc, and −0.24 V for MnPc, with the content increasing accordingly. For IrPc, RhPc, FePc, and CrPc, byproducts-I dominate across the entire potential range. In contrast, byproducts-II exceeds byproducts-I in RuPc between 0 and −0.4 V, and in MnPc below −0.15 V, primarily due to the accumulation of $^M$O–O$_{outer}$ (**Figure 4g**). At a representative potential of −0.1 V *vs.* SHE, the total byproduct content follows the order of IrPc > MnPc > RhPc > RuPc > CrPc > FePc.



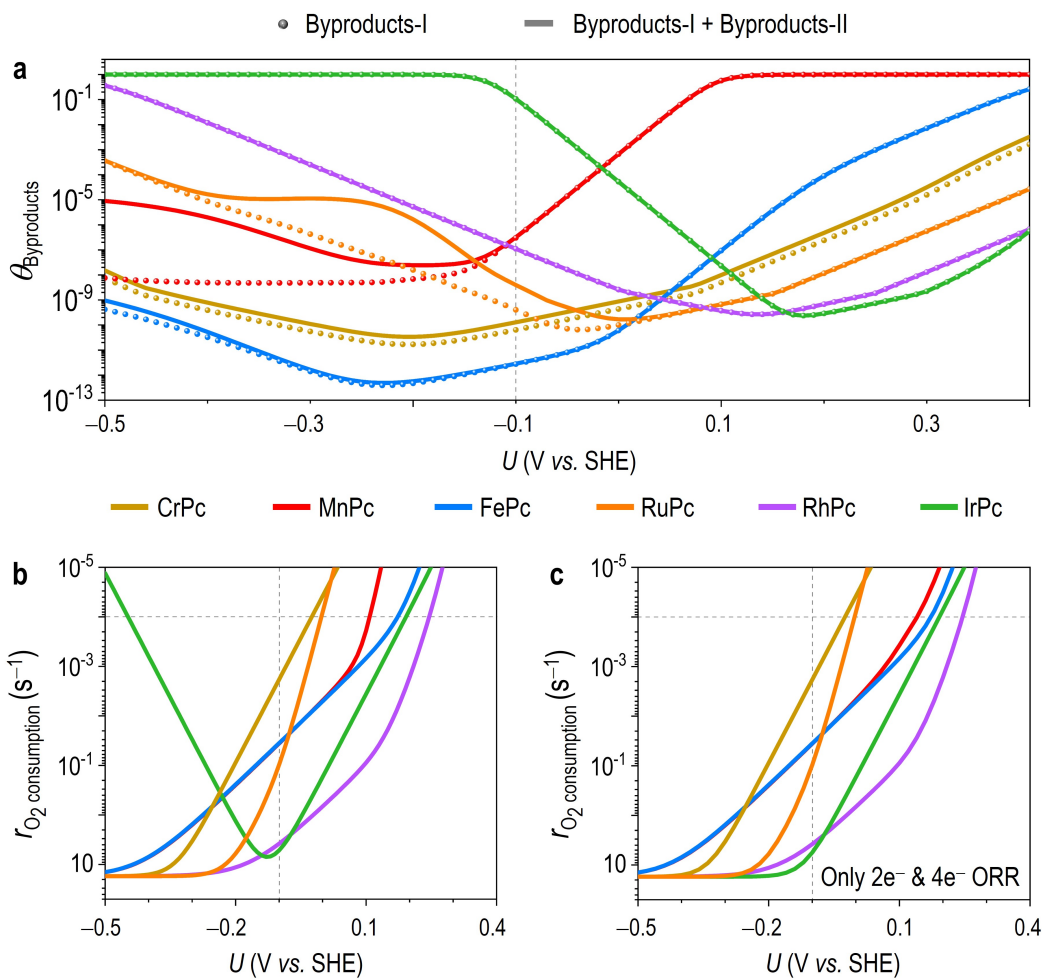

**Figure 6.** Potential-dependent evolution of the total content of (a) products; (b) oxygen consumption rate associated complete degradation networks; (c) $2e^-$ and $4e^-$ ORR pathways for all examined MPcs at pH 13 after 3600 s.

Regarding the potential-dependent $O_2$ consumption rates (**Figure 6b**), the onset potential, defined here as potential at which the rate reaches $10^{-4}$ $s^{-1}$, follows the trend of RhPc (0.25 V) > IrPc (0.19 V) > FePc (0.17 V) > MnPc (0.11 V) > RuPc (0.00 V) > CrPc (−0.03 V). Previous experiments indeed find that the ORR activity of FePc is greater than MnPc[5]. Compared to the results without accounting for the degradation network (**Figure 6c**), two major differences are observed. First, the onset potential for MnPc decreases slightly (0.11 V in **Figure 6b** vs. 0.14 V in **Figure 6c**), which is attributed to the formation of $^{M}OH–^{Cl}OH$. Second, IrPc exhibits a sharp drop in $O_2$ consumption rate below −0.15 V *vs.* SHE, due to the accumulation of the $^{M}H$ state (**Figure 4b**), which blocks the metal center and hinders ORR activity.

## 4. CONCLUSIONS

In summary, we integrate first-principles calculations of ORR pathways, ROS formation, and byproduct generation with time-dependent microkinetic modeling to quantitively investigate the degradation mechanisms of MPc catalysts during ORR. Our findings reveal that the major byproducts arise from carbon oxidation, primarily driven by •OH radical attack and structural reorganization of adsorbates, as



well as from protonation at metal center or the N2 site. Demetallation is negligible due to the high energy barriers for metal ion displacement and the difficulty of protonation at the N1 site. Under the examined operating conditions, MnPc and FePc primarily undergo carbon oxidation. CrPc and RuPc also exhibit dominant carbon oxidation, but are additionally susceptible to protonation at the N2 site under low potential and low pH conditions. In contrast, RhPc and IrPc mainly undergo protonation at the metal center, with partial susceptibility to carbon oxidation at high potential and high pH. In the kinetics-controlled region, the ORR activity ranks as RhPc > IrPc > FePc > MnPc > RuPc > CrPc. These findings suggest that selecting a proper metal center, reinforcing the vulnerable C1 sites against oxidation, suppressing ROS generation, and optimizing the operating potential are promising strategies to enhance the long-term stability of MPcs catalysts during ORR.

## ASSOCIATED CONTENT

### Supplementary Information

The Supporting information includes the literature survey on the stability of M-N-C catalysts under alkaline and acidic ORR conditions; free energy diagrams for the 4e⁻ and 2e⁻ ORR processes on MPcs; microkinetic model with only the 4e⁻ and 2e⁻ ORR processes; comparison of adsorption ability at different sites of FePc; comparison of protonation at different sites of MPcs; radical attacks on the metal and C1 sites of MPcs; generation and conversion of ROS in solution and on MPcs; microkinetic model with ORR processes and side reactions; comparison between direct metal-leaching pathway and oxidation-induced demetallation processes; variation of major byproduct distribution with pH and potential; reaction network of each MPc under typical experimental conditions; and Mathematica code for the time-dependent microkinetic model.

## AUTHOR INFORMATION


### Corresponding Author
*E-mail: luogf@sustech.edu.cn

### Author Contributions
G.L. conceived and oversaw the project. H.Y. performed the calculations and data collection. G.L. provided guidance on the study design and interpretation of results. G.L. and H.Y. analyzed the data and wrote the manuscript.


### Notes
The authors declare that they have no known competing financial interests or personal relationships that could have appeared to influence the work reported in this paper.

## ACKNOWLEDGEMENTS




This work was supported by the fund of the National Foundation of Natural Science, China (No. 52273226), the Guangdong Provincial Key Laboratory of Computational Science and Material Design (No. 2019B030301001), the Guangdong Basic and Applied Basic Research Foundation (No. 2024A1515010211), the Shenzhen Science and Technology Innovation Commission (No. JCYJ20200109141412308), and the High level of special funds of Southern University of Science and Technology (No. G03050K002). All the calculations were carried out on the Taiyi cluster supported by the Center for Computational Science and Engineering of the Southern University of Science and Technology.



**REFERENCES**

(1) Kumar, A.; Zhang, Y.; Liu, W.; Sun, X. The chemistry, recent advancements and activity descriptors for macrocycles based electrocatalysts in oxygen reduction reaction. *Coord. Chem. Rev.* **2020**, *402*, 213047. DOI: 10.1016/j.ccr.2019.213047.

(2) Debe, M. K. Electrocatalyst approaches and challenges for automotive fuel cells. *Nature* **2012**, *486* (7401), 43-51. DOI: 10.1038/nature11115.

(3) Tong, M.; Wang, L.; Fu, H. Designed synthesis and catalytic mechanisms of non-precious metal single-atom catalysts for oxygen reduction reaction. *Small Methods* **2021**, *5* (10), e2100865. DOI: 10.1002/smtd.202100865.

(4) Jiang, Y. Y.; Lu, Y. Z.; Lv, X. Y.; Han, D. X.; Zhang, Q. X.; Niu, L.; Chen, W. Enhanced catalytic performance of Pt-free iron phthalocyanine by graphene support for efficient oxygen reduction reaction. *ACS Catal.* **2013**, *3* (6), 1263-1271. DOI: 10.1021/cs4001927.

(5) Singh, D. K.; Ganesan, V.; Yadav, D. K.; Yadav, M. Metal (Mn, Fe, Co, Ni, Cu, and Zn) phthalocyanine-immobilized mesoporous carbon nitride materials as durable electrode modifiers for the oxygen reduction reaction. *Langmuir* **2020**, *36* (41), 12202-12212. DOI: 10.1021/acs.langmuir.0c01822.

(6) Guo, Y.; Yin, H.; Cheng, F.; Li, M.; Zhang, S.; Wu, D.; Wang, K.; Wu, Y.; Yang, B.; Zhang, J. N. Altering ligand microenvironment of atomically dispersed $CrN_4$ by axial ligand sulfur for enhanced oxygen reduction reaction in alkaline and acidic medium. *Small* **2023**, *19* (14), 2206861. DOI: 10.1002/smll.202206861.

(7) Li, J.; Chen, M.; Cullen, D. A.; Hwang, S.; Wang, M.; Li, B.; Liu, K.; Karakalos, S.; Lucero, M.; Zhang, H.; et al. Atomically dispersed manganese catalysts for oxygen reduction in proton-exchange membrane fuel cells. *Nat. Catal.* **2018**, *1* (12), 935-945. DOI: 10.1038/s41929-018-0164-8.

(8) Li, J.; Chen, S.; Yang, N.; Deng, M.; Ibraheem, S.; Deng, J.; Li, J.; Li, L.; Wei, Z. Ultrahigh-loading zinc single-atom catalyst for highly efficient oxygen reduction in both acidic and alkaline media. *Angew. Chem. Int. Ed.* **2019**, *58* (21), 7035-7039. DOI: 10.1002/anie.201902109.

(9) Zhang, D.; Wang, Z.; Liu, F.; Yi, P.; Peng, L.; Chen, Y.; Wei, L.; Li, H. Unraveling the pH-dependent oxygen reduction performance on single-atom catalysts: from single- to dual-Sabatier optima. *Journal of the American Chemical Society* **2024**, *146* (5), 3210-3219. DOI: 10.1021/jacs.3c11246.

(10) Luo, E.; Zhang, H.; Wang, X.; Gao, L.; Gong, L.; Zhao, T.; Jin, Z.; Ge, J.; Jiang, Z.; Liu, C.; et al. Single-atom $Cr-N_4$ sites designed for durable oxygen reduction catalysis in acid media. *Angew. Chem. Int. Ed.* **2019**, *58* (36), 12469-12475. DOI: 10.1002/anie.201906289.

(11) Yang, G.; Zhu, J.; Yuan, P.; Hu, Y.; Qu, G.; Lu, B. A.; Xue, X.; Yin, H.; Cheng, W.; Cheng, J.; et al.





Regulating Fe-spin state by atomically dispersed Mn-N in Fe-N-C catalysts with high oxygen reduction activity. *Nat. Commun.* **2021**, *12* (1), 1734. DOI: 10.1038/s41467-021-21919-5.

(12) Xie, X.; He, C.; Li, B.; He, Y.; Cullen, D. A.; Wegener, E. C.; Kropf, A. J.; Martinez, U.; Cheng, Y.; Engelhard, M. H.; et al. Performance enhancement and degradation mechanism identification of a single-atom Co–N–C catalyst for proton exchange membrane fuel cells. *Nat. Catal.* **2020**, *3* (12), 1044-1054. DOI: 10.1038/s41929-020-00546-1.

(13) Xiao, M.; Gao, L.; Wang, Y.; Wang, X.; Zhu, J.; Jin, Z.; Liu, C.; Chen, H.; Li, G.; Ge, J.; et al. Engineering energy level of metal center: Ru single-atom site for efficient and durable oxygen reduction catalysis. *J. Am. Chem. Soc.* **2019**, *141* (50), 19800-19806. DOI: 10.1021/jacs.9b09234.

(14) Wen, X.; Yu, C.; Yan, B.; Zhang, X.; Liu, B.; Xie, H.; Kang Shen, P.; Qun Tian, Z. Morphological and microstructural engineering of Mn-N-C with strengthened Mn-N bond for efficient electrochemical oxygen reduction reaction. *Chem. Eng. J.* **2023**, *475*, 146135. DOI: 10.1016/j.cej.2023.146135.

(15) Gao, H.; Wang, Y.; Li, W.; Zhou, S.; Song, S.; Tian, X.; Yuan, Y.; Zhou, Y.; Zang, J. One-step carbonization of ZIF-8 in Mn-containing ambience to prepare Mn, N co-doped porous carbon as efficient oxygen reduction reaction electrocatalyst. *Int. J. Hydrogen Energy* **2021**, *46* (74), 36742-36752. DOI: 10.1016/j.ijhydene.2021.08.220.

(16) Zhang, Z. P.; Dou, M. L.; Ji, J.; Wang, F. Phthalocyanine tethered iron phthalocyanine on graphitized carbon black as superior electrocatalyst for oxygen reduction reaction. *Nano Energy* **2017**, *34*, 338-343. DOI: 10.1016/j.nanoen.2017.02.042.

(17) Yu, X.; Lai, S.; Xin, S.; Chen, S.; Zhang, X.; She, X.; Zhan, T.; Zhao, X.; Yang, D. Coupling of iron phthalocyanine at carbon defect site via π-π stacking for enhanced oxygen reduction reaction. *Appl. Catal. B: Environ.* **2021**, *280*, 119437. DOI: 10.1016/j.apcatb.2020.119437.

(18) Zhang, W.; Wang, L.; Zhang, L. H.; Chen, D.; Zhang, Y.; Yang, D.; Yan, N.; Yu, F. Creating hybrid coordination environment in Fe-based single atom catalyst for efficient oxygen reduction. *ChemSusChem* **2022**, *15* (12), e202200195. DOI: 10.1002/cssc.202200195.

(19) Zhang, W.; Meeus, E. J.; Wang, L.; Zhang, L. H.; Yang, S.; de Bruin, B.; Reek, J. N. H.; Yu, F. Boosting electrochemical oxygen reduction performance of iron phthalocyanine through axial coordination sphere interaction. *ChemSusChem* **2022**, *15* (3), e202102379. DOI: 10.1002/cssc.202102379.

(20) Liu, D.; Long, Y.-T. Superior catalytic activity of electrochemically reduced graphene oxide supported iron phthalocyanines toward oxygen reduction reaction. *ACS Appl. Mater. Inter.* **2015**, *7* (43), 24063-24068. DOI: 10.1021/acsami.5b07068.

(21) Zhang, X.; Chen, C.; Dong, J.; Wang, R. X.; Wang, Q.; Zhou, Z. Y.; Sun, S. G. Comparative study of the oxygen reduction reaction on pyrolyzed FePc in acidic and alkaline Media. *ChemElectroChem* **2018**, *5* (24), 3946-3952. DOI: 10.1002/celc.201801179.

(22) Bae, G.; Chung, M. W.; Ji, S. G.; Jaouen, F.; Choi, C. H. pH effect on the $H_2O_2$-induced deactivation of Fe-N-C catalysts. *ACS Catal.* **2020**, *10* (15), 8485-8495. DOI: 10.1021/acscatal.0c00948.

(23) Xie, H.; Xie, X.; Hu, G.; Prabhakaran, V.; Saha, S.; Gonzalez-Lopez, L.; Phakatkar, A. H.; Hong, M.; Wu, M.; Shahbazian-Yassar, R.; et al. Ta–TiO$_x$ nanoparticles as radical scavengers to improve the durability of Fe–N–C oxygen reduction catalysts. *Nat. Energy* **2022**, *7* (3), 281-289. DOI: 10.1038/s41560-022-00988-w.

(24) Chen, J.; Yan, X.; Fu, C.; Feng, Y.; Lin, C.; Li, X.; Shen, S.; Ke, C.; Zhang, J. Insight into the rapid degradation behavior of nonprecious metal Fe−N−C electrocatalyst-based proton exchange membrane fuel





cells. *ACS Appl. Mater. Interfaces* **2019**, *11* (41), 37779-37786. DOI: 10.1021/acsami.9b13474.

(25) Yang, N.; Peng, L.; Li, L.; Li, J.; Liao, Q.; Shao, M.; Wei, Z. Theoretically probing the possible degradation mechanisms of an FeNC catalyst during the oxygen reduction reaction. *Chem. Sci.* **2021**, *12* (37), 12476-12484. DOI: 10.1039/d1sc02901k.

(26) Baranton, S.; Coutanceau, C.; Roux, C.; Hahn, F.; Leger, J. M. Oxygen reduction reaction in acid medium at iron phthalocyanine dispersed on high surface area carbon substrate: Tolerance to methanol, stability and kinetics. *J. Electroanal. Chem.* **2005**, *577* (2), 223-234. DOI: 10.1016/j.jelechem.2004.11.034.

(27) Herranz, J.; Jaouen, F.; Lefèvre, M.; Kramm, U. I.; Proietti, E.; Dodelet, J.-P.; Bogdanoff, P.; Fiechter, S.; Abs-Wurmbach, I.; Bertrand, P.; et al. Unveiling N-protonation and anion-binding effects on Fe/N/C catalysts for $O_2$ reduction in proton-exchange-membrane fuel cells. *J. Phys. Chem. C* **2011**, *115* (32), 16087-16097. DOI: 10.1021/jp2042526.

(28) Li, H.; Di, S.; Niu, P.; Wang, S.; Wang, J.; Li, L. A durable half-metallic diatomic catalyst for efficient oxygen reduction. *Energy Environ. Sci.* **2022**, *15* (4), 1601-1610. DOI: 10.1039/d1ee03194e.

(29) Yang, N.; Peng, L.; Li, L.; Li, J.; Wei, Z. Theoretical research on the oxidation mechanism of doped carbon based catalysts for oxygen reduction reaction. *Phys. Chem. Chem. Phys.* **2019**, *21* (47), 26102-26110. DOI: 10.1039/c9cp04691g.

(30) Choi, C. H.; Lim, H.-K.; Chung, M. W.; Chon, G.; Ranjbar Sahraie, N.; Altin, A.; Sougrati, M.-T.; Stievano, L.; Oh, H. S.; Park, E. S.; et al. The achilles' heel of iron-based catalysts during oxygen reduction in an acidic medium. *Energy Environ. Sci.* **2018**, *11* (11), 3176-3182. DOI: 10.1039/c8ee01855c.

(31) Tan, X.; Tahini, H. A.; Smith, S. C. Unveiling the role of carbon oxidation in irreversible degradation of atomically-dispersed $FeN_4$ moieties for proton exchange membrane fuel cells. *J. Mater. Chem. A* **2021**, *9* (13), 8721-8729. DOI: 10.1039/d0ta12105c.

(32) Unsal, S.; Girod, R.; Appel, C.; Karpov, D.; Mermoux, M.; Maillard, F.; Saveleva, V. A.; Tileli, V.; Schmidt, T. J.; Herranz, J. Decoupling the contributions of different instability mechanisms to the PEMFC performance decay of non-noble metal $O_2$-reduction catalysts. *J. Am. Chem. Soc.* **2023**, *145* (14), 7845-7858. DOI: 10.1021/jacs.2c12751.

(33) Wan, L.; Zhao, K.; Wang, Y.-C.; Wei, N.; Zhang, P.; Yuan, J.; Zhou, Z.; Sun, S.-G. Molecular degradation of iron phthalocyanine during the oxygen reduction reaction in acidic media. *ACS Catal.* **2022**, *12* (18), 11097-11107. DOI: 10.1021/acscatal.2c03216.

(34) Haber, F.; Weiss, J. On the catalysis of hydroperoxide. *Die Naturwissenschaften* **1932**, *20* (51), 948-950. DOI: 10.1007/bf01504715.

(35) Fritz Haber, J. W. The catalytic decomposition of hydrogen peroxide by iron salts. *Proc. R. Soc. Lond. A* **1934**, *147* (861), 332-351. DOI: 10.1098/rspa.1934.0221.

(36) Brillas, E.; Sires, I.; Oturan, M. A. Electro-Fenton process and related electrochemical technologies based on Fenton's reaction chemistry. *Chem. Rev.* **2009**, *109* (12), 6570-6631. DOI: 10.1021/cr900136g.

(37) Miao, Z.; Li, S.; Priest, C.; Wang, T.; Wu, G.; Li, Q. Effective approaches for designing stable $M–N_x/C$ oxygen-reduction catalysts for proton-exchange membrane fuel cells. *Adv. Mater.* **2022**, *34* (52), 2200595. DOI: 10.1002/adma.202200595.

(38) Yang, W.; Zhao, M.; Ding, X.; Ma, K.; Wu, C.; Gates, I. D.; Gao, Z. The effect of coordination environment on the kinetic and thermodynamic stability of single-atom iron catalysts. *Phys. Chem. Chem. Phys.* **2020**, *22* (7), 3983-3989. DOI: 10.1039/c9cp05349b.




(39) Zhou, Y.; Gao, G.; Chu, W.; Wang, L.-W. Computational screening of transition metal-doped phthalocyanine monolayers for oxygen evolution and reduction. *Nanoscale Adv.* **2020**, *2* (2), 710-716. DOI: 10.1039/c9na00648f.

(40) Di Liberto, G.; Giordano, L.; Pacchioni, G. Predicting the stability of single-atom catalysts in electrochemical reactions. *ACS Catal.* **2023**, *14* (1), 45-55. DOI: 10.1021/acscatal.3c04801.

(41) Orellana, W.; Zuñiga, C.; Gatica, A.; Ureta-Zanartu, M.-S.; Zagal, J. H.; Tasca, F. Effect of electrolyte media on the catalysis of Fe phthalocyanine toward the oxygen reduction reaction: Ab initio molecular dynamics simulations and experimental analyses. *ACS Catal.* **2022**, *12* (20), 12786-12799. DOI: 10.1021/acscatal.2c03298.

(42) Chen, L.-N.; Yu, W.-S.; Wang, T.; Yang, X.-D.; Yang, H.-J.; Chen, Z.-X.; Wang, T.; Tian, N.; Zhou, Z.-Y.; Sun, S.-G. Fluorescence detection of hydroxyl radical generated from oxygen reduction on Fe/N/C catalyst. *Sci. China Chem.* **2019**, *63* (2), 198-202. DOI: 10.1007/s11426-019-9635-2.

(43) Guan, M. H.; Dong, L. Y.; Wu, T.; Li, W. C.; Hao, G. P.; Lu, A. H. Boosting selective oxidation of ethylene to ethylene glycol assisted by in situ generated $H_2O_2$ from $O_2$ electroreduction. *Angew. Chem. Int. Ed.* **2023**, *62* (19), e202302466. DOI: 10.1002/anie.202302466.

(44) *Gaussian 16, Revision C.01*; Gaussian, Inc.: Wallingford, CT, 2016.

(45) Lee, C.; Yang, W.; Parr, R. G. Development of the colle-salvetti correlation-energy formula into a functional of the electron density. *Phys. Rev. B* **1988**, *37* (2), 785-789. DOI: 10.1103/physrevb.37.785.

(46) Becke, A. D. Density-functional exchange-energy approximation with correct asymptotic behavior. *Phys. Rev. A* **1988**, *38* (6), 3098-3100. DOI: 10.1103/physreva.38.3098.

(47) Grimme, S.; Ehrlich, S.; Goerigk, L. Effect of the damping function in dispersion corrected density functional theory. *J. Comput. Chem.* **2011**, *32* (7), 1456-1465. DOI: 10.1002/jcc.21759.

(48) Barone, V.; Cossi, M. Quantum calculation of molecular energies and energy gradients in solution by a conductor solvent model. *J. Phys. Chem. A* **1998**, *102* (11), 1995-2001. DOI: 10.1021/jp9716997.

(49) Schlegel, H. B. Optimization of equilibrium geometries and transition structures. *J. Comput. Chem.* **2004**, *3* (2), 214-218. DOI: 10.1002/jcc.540030212.

(50) Levell, Z.; Le, J.; Yu, S.; Wang, R.; Ethirajan, S.; Rana, R.; Kulkarni, A.; Resasco, J.; Lu, D.; Cheng, J.; et al. Emerging atomistic modeling methods for heterogeneous electrocatalysis. *Chem. Rev.* **2024**, *124* (14), 8620-8656. DOI: 10.1021/acs.chemrev.3c00735.

(51) Abidi, N.; Lim, K. R. G.; Seh, Z. W.; Steinmann, S. N. Atomistic modeling of electrocatalysis: Are we there yet? *WIREs Comput. Mol. Sci.* **2020**, *11* (3), e1499. DOI: 10.1002/wcms.1499.

(52) Mukherjee, B. Investigation of FePc nanoribbon as ORR catalyst in alkaline medium: A DFT based approach. *J. Electrochem. Soc.* **2018**, *165* (15), J3231-J3235. DOI: 10.1149/2.291815jes.

(53) Li, Z.; Ji, S.; Xu, C.; Leng, L.; Liu, H.; Horton, J. H.; Du, L.; Gao, J.; He, C.; Qi, X.; et al. Engineering the electronic structure of single-atom iron sites with boosted oxygen bifunctional activity for zinc–air batteries. *Adv. Mater.* **2022**, *35* (9), 2209644. DOI: 10.1002/adma.202209644.

(54) Zhang, H.; Zhang, S.; Wang, Y.; Si, J.; Chen, Y.; Zhuang, L.; Chen, S. Boosting the performance of iron-phthalocyanine as cathode electrocatalyst for alkaline polymer fuel cells through edge-closed conjugation. *ACS Appl. Mater. Interfaces* **2018**, *10* (34), 28664-28671. DOI: 10.1021/acsami.8b09074.

(55) Medford, A. J.; Shi, C.; Hoffmann, M. J.; Lausche, A. C.; Fitzgibbon, S. R.; Bligaard, T.; Nørskov, J. K.




CatMAP: A software package for descriptor-based microkinetic mapping of catalytic trends. *Catal. Lett.* **2015**, *145* (3), 794-807. DOI: 10.1007/s10562-015-1495-6.

(56) Carneiro-Neto, E. B.; Lopes, M. C.; Pereira, E. C. Simulation of interfacial pH changes during hydrogen evolution reaction. *J. Electroanal. Chem.* **2016**, *765*, 92-99. DOI: 10.1016/j.jelechem.2015.09.029.

(57) Huang, J.; Zhang, J.; Eikerling, M. Unifying theoretical framework for deciphering the oxygen reduction reaction on platinum. *Phys. Chem. Chem. Phys.* **2018**, *20* (17), 11776-11786. DOI: 10.1039/c8cp01315b.

(58) Yang, H. H.; Bashir, B.; Luo, G. F. Towards superior metal phthalocyanine catalysts for electrochemical oxygen reduction: A comprehensive screening under experimental conditions. *Chem. Eng. J.* **2023**, *473*, 145101. DOI: 10.1016/j.cej.2023.145101.

(59) Wang, Y.; Yuan, H.; Li, Y. F.; Chen, Z. F. Two-dimensional iron-phthalocyanine (Fe-Pc) monolayer as a promising single-atom-catalyst for oxygen reduction reaction: A computational study. *Nanoscale* **2015**, *7* (27), 11633-11641. DOI: 10.1039/c5nr00302d.

(60) Schumb, W.; Satterfield, C. N.; Wentworth, R. L. *Hydrogen peroxide*; Reinhold Publishing Corporation, 1955. DOI: 10.1002/jps.3030450224.

(61) Zhang, T.; Liu, Y.; Wang, Y.; Wang, Z.; Liu, J.; Gong, X. Generation and transfer of long lifetime reactive oxygen species (ROSs) from electrochemical regulation. *Chem. Eng. J.* **2023**, *464*, 142443. DOI: 10.1016/j.cej.2023.142443.

(62) Danilczuk, M.; Coms, F. D.; Schlick, S. Visualizing chemical reactions and crossover processes in a fuel cell inserted in the ESR resonator: Detection by spin trapping of oxygen radicals, nafion-derived fragments, and hydrogen and deuterium atoms. *J. Phys. Chem. B* **2009**, *113* (23), 8031-8042. DOI: 10.1021/jp901597f.

(63) Pandis, S. N.; Seinfeld, J. H. Sensitivity analysis of a chemical mechanism for aqueous-phase atmospheric chemistry. *J. Geophys. Res.: Atmos.* **2012**, *94* (D1), 1105-1126. DOI: 10.1029/JD094iD01p01105.

(64) Zeng, H.; Zhang, G.; Ji, Q.; Liu, H.; Hua, X.; Xia, H.; Sillanpaa, M.; Qu, J. pH-independent production of hydroxyl radical from atomic H*-mediated electrocatalytic $H_2O_2$ reduction: A green Fenton process without byproducts. *Environ. Sci. Technol.* **2020**, *54* (22), 14725-14731. DOI: 10.1021/acs.est.0c04694.

(65) Duan, Z.; Henkelman, G. Surface charge and electrostatic spin crossover effects in $CoN_4$ electrocatalysts. *ACS Catal.* **2020**, *10* (20), 12148-12155. DOI: 10.1021/acscatal.0c02458.




**Supporting information**

# Atomistic Insights into the Degradation of Metal Phthalocyanines Catalysts during Oxygen Reduction Reaction


Huanhuan Yang[1,2,3] and Guangfu Luo[2,3,4,*]

[1]Harbin Institute of Technology, Harbin 150080, P.R. China

[2]State Key Laboratory of Quantum Functional Materials, Department of Materials Science and Engineering, Southern University of Science and Technology, Shenzhen 518055, P. R. China

[3]Guangdong Provincial Key Laboratory of Computational Science and Material Design, Southern University of Science and Technology, Shenzhen 518055, P.R. China

[4]Institute of Innovative Materials, Southern University of Science and Technology, Shenzhen 518055, P.R. China

*E-mail: luogf@sustech.edu.cn


## Contents





# 1. Literature survey on the stability of M-N-C for ORR under alkaline and acidic conditions

**Table S1.** Summary of literature data on the initial half-wave potential ($E_{1/2}$) and stability of M-N-C catalysts for ORR under alkaline (green) and acidic (orange) conditions. Stability is characterized by the relative change of current density, $(i_0 - i)/i_0$, the change of half-wave potential ($\Delta E_{1/2}$), or average loss rate. Performances of catalysts Pt/C are also provided for comparison.

| Material | pH | Initial $E_{1/2}$ (V vs. RHE) | $(i_0-i)/i_0$ or $\Delta E_{1/2}$ (optional condition) | Total loss/Total time | Year | Ref. |
|---|---|---|---|---|---|---|
| FePc/DG | 13 | 0.90 | 4% (15 h at 0.75 V) | 0.27% h$^{-1}$ | 2021 | 1 |
| Pt/C | | 0.86 | 13% (15 h at 0.75 V) | 0.87% h$^{-1}$ | | |
| FePc/SNGO | 13 | 0.94 | 9% (6.94 h at 0.7 V)<br>15 mV (5×10$^3$ cycles) | 1.30% h$^{-1}$ | 2022 | 2 |
| Pt/C | | 0.85 | 19% (6.94 h at 0.7 V)<br>53 mV (5×10$^3$ cycles) | 2.74% h$^{-1}$ | | |
| FePc/CNT | 13 | 0.85 | 10% (6.94 h at unknown voltage) | 1.44% h$^{-1}$ | 2022 | 3 |
| Pt/C | | 0.85 | 19% (6.94 h at unknown voltage) | 2.74% h$^{-1}$ | | |
| FePc/ERGO | 13 | 0.92 | 12% (2.22 h at 0.75 V) | 5.41% h$^{-1}$ | 2015 | 4 |
| Pt/C | | 0.87 | 38% (2.22 h at 0.75 V) | 17.1% h$^{-1}$ | | |
| FePc/C | 13 | 0.86 | 53.3% (2.78 h at 0.9 V) | 19.2% h$^{-1}$ | 2017 | 5 |
| Pt/C | | 0.84 | 37.4% (2.78 h at 0.9 V) | 13.5% h$^{-1}$ | | |
| MnPc | 13 | 0.54 | 20 mV (1.5×10$^3$ cycles) | Not provided | 2019 | 6 |
| Fe-N-C-1 | 13 | 0.88 | 18.67 mV (10$^3$ cycles, 2.22 h) | 8.41 mV h$^{-1}$ | 2019 | 7 |
| Zn-N-C-1 | | 0.87 | 0.54 mV (10$^3$ cycles, 2.22 h) | 0.24 mV h$^{-1}$ | | |
| Fe-N-C-1 | 1 | 0.74 | 31 mV (10$^3$ cycles, 2.22 h) | 13.96 mV h$^{-1}$ | | |
| Zn-N-C-1 | | 0.75 | 19 mV (10$^3$ cycles, 2.22 h) | 8.56 mV h$^{-1}$ | | |
| Mn-N-C | 13 | 0.88 | 11 mV (3×10$^4$ cycles, 33.33 h) | 0.33 mV h$^{-1}$ | 2023 | 8 |
| Pt/C | | 0.86 | 26 mV (3×10$^4$ cycles, 33.33 h) | 0.78 mV h$^{-1}$ | | |
| Mn-N-C$^a$ | 1 | 0.73 | 34 mV (3×10$^4$ cycles, 33.33 h) | 1.02 mV h$^{-1}$ | | |
| Pt/C$^a$ | | 0.82 | 37 mV (3×10$^4$ cycles, 33.33 h) | 1.11 mV h$^{-1}$ | | |
| Mn-N-C | 13 | 0.88 | 6.6% (5 h at 0.7 V) | 1.32% h$^{-1}$ | 2021 | 9 |
| Pt/C | | 0.84 | 20.9% (5 h at 0.7 V) | 4.18% h$^{-1}$ | | |
| Mn-N-C$^a$ | 1 | 0.73 | 22 mV (5×10$^3$ cycles, 11.11 h)<br>7.9% (5 h at 0.6V) | 1.98 mV h$^{-1}$<br>1.58% h$^{-1}$ | | |
| Pt/C$^a$ | | 0.78 | 38 mV (5×10$^3$ cycles, 11.11 h) | 3.60 mV h$^{-1}$ | | |
| Mn-N-C | 13 | 0.88 | 16 mV (10$^4$ cycles, 11.11 h) | 1.44 mV h$^{-1}$ | 2020 | 10 |
| Pt/C | | 0.86 | 34 mV (10$^4$ cycles, 11.11 h) | 3.06 mV h$^{-1}$ | | |



[a] 0.5 M $H_2SO_4$

DG: defective graphene

SNGO: N, S co-doped graphene oxide

CNT: multi-walled carbon nanotubes

ERGO: electrochemically reduced graphene oxide



## 2. Microkinetic model with ORR processes and side reactions

**Table S2.** Elementary reaction steps and net reaction rates for ORR processes and side reactions. The Gibbs free energy change and forward barrier for each step is provided in Supplementary Materials as a separate file, **energy.xslx**.

| Label | Elementary Reaction Step | Net Reaction Rate ($r_i$) |
|---|---|---|
| | 4e$^-$ and 2e$^-$ ORR processes | |
| 1 | $O_2\,(aq) \rightarrow O_2\,(dl)$ | $8.0 \times 10^5 \times (x_{O_2,\,aq} - x_{O_2,\,dl})$ |
| 2 | $M + O_2\,(dl) \rightarrow {}^M O_2$ | $f_2 \times \theta^M \times x_{O_2,\,dl} - b_2 \times \theta^{MO_2}$ |
| 3 | ${}^M O_2 + H^+ + e^- \rightarrow {}^M OOH$ | $f_3 \times \theta^{MO_2} - b_3 \times \theta^{MOOH}$ |
| 4 | ${}^M OOH + H^+ + e^- \rightarrow {}^M O + H_2O\,(l)$ | $f_4 \times \theta^{MOOH} - b_4 \times \theta^{MO}$ |
| 5 | ${}^M O + H^+ + e^- \rightarrow {}^M OH$ | $f_5 \times \theta^{MO} - b_5 \times \theta^{MOH}$ |
| 6 | ${}^M OH + H^+ + e^- \rightarrow M + H_2O\,(l)$ | $f_6 \times \theta^{MOH} - b_6 \times \theta^M$ |
| 7 | ${}^M OOH + H^+ + e^- \rightarrow {}^M H_2O_2$ | $f_7 \times \theta^{MOOH} - b_7 \times \theta^{MH_2O_2}$ |
| 8 | ${}^M H_2O_2 \rightarrow M + H_2O_2\,(aq)$ | $f_8 \times \theta^{MH_2O_2} - b_8 \times \theta^M \times x(H_2O_2)$ |
| | Protonation for the generation of •OH radical | |
| 9 | ${}^M O_2 + H^+ + e^- \rightarrow {}^M O_2\text{–}^{N2}H$ | $f_9 \times \theta^{MO_2} - b_9 \times \theta^{MO_2\text{–}N2H}$ |
| 10 | ${}^M O_2\text{–}^{N2}H + H_2O_2\,(aq) \rightarrow {}^M O_2 + \bullet OH + H_2O\,(l)$ | $f_{10} \times \theta^{MO_2\text{–}N2H} \times x(H_2O_2) - b_{10} \times \theta^{MO_2} \times x(\bullet OH)$ |
| 11 | ${}^M OOH + H^+ + e^- \rightarrow {}^M OOH\text{–}^{N2}H$ | $f_{11} \times \theta^{MOOH} - b_{11} \times \theta^{MOOH\text{–}N2H}$ |
| 12 | ${}^M OOH\text{–}^{N2}H + H_2O_2\,(aq) \rightarrow {}^M OOH + \bullet OH + H_2O\,(l)$ | $f_{12} \times \theta^{MOOH\text{–}N2H} \times x(H_2O_2) - b_{12} \times \theta^{MOOH} \times x(\bullet OH)$ |
| 13 | ${}^M O + H^+ + e^- \rightarrow {}^M O\text{–}^{N2}H$ | $f_{13} \times \theta^{MO} - b_{13} \times \theta^{MO\text{–}N2H}$ |
| 14 | ${}^M O\text{–}^{N2}H + H_2O_2\,(aq) \rightarrow {}^M O + \bullet OH + H_2O\,(l)$ | $f_{14} \times \theta^{MO\text{–}N2H} \times x(H_2O_2) - b_{14} \times \theta^{MO} \times x(\bullet OH)$ |
| 15 | ${}^M OH + H^+ + e^- \rightarrow {}^M OH\text{–}^{N2}H$ | $f_{15} \times \theta^{MOH} - b_{15} \times \theta^{MOH\text{–}N2H}$ |
| 16 | ${}^M OH\text{–}^{N2}H + H_2O_2\,(aq) \rightarrow {}^M OH + \bullet OH + H_2O\,(l)$ | $f_{16} \times \theta^{MOH\text{–}N2H} \times x(H_2O_2) - b_{16} \times \theta^{MOH} \times x(\bullet OH)$ |
| 17 | $M + H^+ + e^- \rightarrow {}^M H$ | $f_{17} \times \theta^M - b_{17} \times \theta^{MH}$ |
| 18 | ${}^M H + H_2O_2\,(aq) \rightarrow M + \bullet OH + H_2O\,(l)$ | $f_{18} \times \theta^{MH} \times x(H_2O_2) - b_{18} \times \theta^M \times x(\bullet OH)$ |
| 19 | $M + H^+ + e^- \rightarrow {}^{N2}H$ | $f_{19} \times \theta^M - b_{19} \times \theta^{N2H}$ |
| 20 | ${}^{N2}H + H_2O_2\,(aq) \rightarrow M + \bullet OH + H_2O\,(l)$ | $f_{20} \times \theta^{N2H} \times x(H_2O_2) - b_{20} \times \theta^M \times x(\bullet OH)$ |
| 21 | ${}^M H_2O_2 \rightarrow {}^M OH + \bullet OH$ | $f_{21} \times \theta^{MH_2O_2} - b_{21} \times \theta^{MOH} \times x(\bullet OH)$ |
| 22 | ${}^M OOH \rightarrow {}^M O + \bullet OH$ | $f_{22} \times \theta^{MOOH} - b_{22} \times \theta^{MO} \times x(\bullet OH)$ |
| | Degradation processes involving radical and protonation, facilitating interconversion among adsorbates | |



(*Continued*) **Table S2.**

| Label | Elementary Reaction Step | Net Reaction Rate ($r_i$) |
|---|---|---|
| 23 | $^M$OOH → $^M$O–$^{C1}$OH | $f_{23} \times \theta^{MOOH} - b_{23} \times \theta^{MO-C1OH}$ |
| 24 | M + •OOH → $^M$OOH | $f_{24} \times \theta^M \times x(\text{•OOH}) - b_{24} \times \theta^{MOOH}$ |
| 25 | M + •OH → $^M$OH | $f_{25} \times \theta^M \times x(\text{•OH}) - b_{25} \times \theta^{MOH}$ |
| 26 | $^M$OH + •OOH → $^M$OH–$^{C1}$OOH | $f_{26} \times \theta^{MOH} \times x(\text{•OOH}) - b_{26} \times \theta^{MOH-C1OOH}$ |
| 27 | $^M$OH + •OH → $^M$OH–$^{C1}$OH | $f_{27} \times \theta^{MOH} \times x(\text{•OH}) - b_{27} \times \theta^{MOH-C1OH}$ |
| 28 | $^M$OH + •OH → $^M$2OH | $f_{28} \times \theta^{MOH} \times x(\text{•OH}) - b_{28} \times \theta^{M2OH}$ |
| 29 | $^M$2OH + •OH → $^M$3OH | $f_{29} \times \theta^{M2OH} \times x(\text{•OH}) - b_{29} \times \theta^{M3OH}$ |
| 30 | $^M$O + •OOH → $^M$O–$^{C1}$OOH | $f_{30} \times \theta^{MO} \times x(\text{•OOH}) - b_{30} \times \theta^{MO-C1OOH}$ |
| 31 | $^M$O + •OH → $^M$O–$^{C1}$OH | $f_{31} \times \theta^{MO} \times x(\text{•OH}) - b_{31} \times \theta^{MO-C1OH}$ |
| 32 | $^M$O + •OH → $^M$O–$^M$OH | $f_{32} \times \theta^{MO} \times x(\text{•OH}) - b_{32} \times \theta^{MO-MOH}$ |
| 33 | $^M$OOH + •OH → $^M$OOH–$^{C1}$OH | $f_{33} \times \theta^{MOOH} \times x(\text{•OH}) - b_{33} \times \theta^{MOOH-C1OH}$ |
| 34 | $^M$OOH–$^{C1}$OH → $^M$H$_2$O$_2$–$^{C1}$O | $f_{34} \times \theta^{MOOH-C1OH} - b_{34} \times \theta^{MH_2O_2-C1O}$ |
| 35 | $^M$H$_2$O$_2$–$^{C1}$O → H$_2$O$_2$ (*aq*) + $^{C1}$O | $f_{35} \times \theta^{MH_2O_2-C1O} - b_{35} \times \theta^{C1O} \times x(H_2O_2)$ |
| 36 | $^M$OH–$^{C1}$OH → $^M$H$_2$O–$^{C1}$O | $f_{36} \times \theta^{MOH-C1OH} - b_{36} \times \theta^{MH_2O-C1O}$ |
| 37 | $^M$H$_2$O–$^{C1}$O → H$_2$O (*l*) + $^{C1}$O | $f_{37} \times \theta^{MH_2O-C1O} - b_{37} \times \theta^{C1O}$ |
| 38 | $^{C1}$O → O$_{inter}$ | $f_{38} \times \theta^{C1O} - b_{38} \times \theta O_{inter}$ |
| 39 | $^M$OOH–$^{C1}$OH → $^M$O–$^{C1}$O + H$_2$O | $f_{39} \times \theta^{MOOH-C1OH} - b_{39} \times \theta^{MO-C1O}$ |
| 40 | $^M$OH–$^{C1}$OOH → $^M$O–$^{C1}$O + H$_2$O | $f_{40} \times \theta^{MOH-C1OOH} - b_{40} \times \theta^{MO-C1O}$ |
| 41 | $^M$O–$^{C1}$O → $^M$O–O$_{outer}$ | $f_{41} \times \theta^{MO-C1O} - b_{41} \times \theta^{MO-C1O}$ |
| 42 | $^M$O–$^{C1}$OH → $^M$OH–$^{C1}$O | $f_{42} \times \theta^{MO-C1OH} - b_{42} \times \theta^{MOH-C1O}$ |
| 43 | $^M$OH–$^{C1}$O → $^M$OH–O$_{inter}$ | $f_{43} \times \theta^{MOH-C1O} - b_{43} \times \theta^{MOH-O_{inter}}$ |
| 44 | $^M$OH–O$_{inter}$ + H$^+$ + e$^-$ → H$_2$O + O$_{inter}$ | $f_{44} \times \theta^{MOH-O_{inter}} - b_{44} \times \theta O_{inter}$ |
| 45 | $^M$O–$^{C1}$OH + H$^+$ + e$^-$ → $^M$OH–$^{C1}$OH | $f_{45} \times \theta^{MO-C1OH} - b_{45} \times \theta^{MOH-C1OH}$ |
| 46 | $^M$O–$^{C1}$OOH + H$^+$ + e$^-$ → $^M$OH–$^{C1}$OOH | $f_{46} \times \theta^{MO-C1OOH} - b_{46} \times \theta^{MOH-C1OOH}$ |
| 47 | $^M$O–$^M$OH + H$^+$ + e$^-$ → $^M$2OH | $f_{47} \times \theta^{MO-MOH} - b_{47} \times \theta^{M2OH}$ |
| 48 | $^M$2OH + H$^+$ + e$^-$ → $^M$2OH–$^{N1}$H | $f_{48} \times \theta^{M2OH} - b_{48} \times \theta^{M2OH-N1H}$ |
| 49 | $^M$2OH–$^{N1}$H + H$^+$ + e$^-$ → $^M$2OH–$^{N1}$H–$^{N3}$H | $f_{49} \times \theta^{M2OH-N1H} - b_{49} \times \theta^{M2OH-N1H-N3H}$ |
| 50 | $^M$2OH–$^{N1}$H–$^{N3}$H → Pc+M(OH)$_2$ | $f_{50} \times \theta^{M2OH-N1H-N3H} - b_{50} \times \theta Pc+M(OH)_2$ |



(*Continued*) **Table S2.**

| Label | Elementary Reaction Step | Net Reaction Rate ($r_i$) |
|---|---|---|
| 51 | $^M3OH + H^+ + e^- \rightarrow {}^M3OH–{}^{N1}H$ | $f_{51} \times \theta^{M}3OH - b_{51} \times \theta^{M}3OH–{}^{N1}H$ |
| 52 | $^M3OH–{}^{N1}H + H^+ + e^- \rightarrow {}^M3OH–{}^{N1}H–{}^{N3}H$ | $f_{52} \times \theta^{M}3OH–{}^{N1}H - b_{52} \times \theta^{M}3OH–{}^{N1}H–{}^{N3}H$ |
| 53 | $^M3OH–{}^{N1}H–{}^{N3}H \rightarrow Pc+M(OH)_3$ | $f_{53} \times \theta^{M}3OH–{}^{N1}H–{}^{N3}H - b_{53} \times \theta Pc+M(OH)_3$ |
| 54 | $M + {}^\bullet OH \rightarrow {}^{C1}OH$ | $f_{54} \times \theta^{M} \times x({}^\bullet OH) - b_{54} \times \theta^{C1}OH$ |
| 55 | $^{C1}OH + O_2\,(dl) \rightarrow {}^{M}O_2–{}^{C1}OH$ | $f_{55} \times \theta^{C1}OH \times x_{O_2,\,dl} - b_{55} \times \theta^{M}O_2–{}^{C1}OH$ |
| 56 | $^{M}O_2–{}^{C1}OH + H^+ + e^- \rightarrow {}^{M}OOH–{}^{C1}OH$ | $f_{56} \times \theta^{M}O_2–{}^{C1}OH - b_{56} \times \theta^{M}OOH–{}^{C1}OH$ |
| 57 | $^{M}OOH–{}^{C1}OH + H^+ + e^- \rightarrow {}^{M}O–{}^{C1}OH + H_2O$ | $f_{57} \times \theta^{M}OOH–{}^{C1}OH - b_{57} \times \theta^{M}O–{}^{C1}OH$ |
| 58 | $^{M}OH–{}^{C1}OH + H^+ + e^- \rightarrow {}^{C1}OH + H_2O$ | $f_{58} \times \theta^{M}OH–{}^{C1}OH - b_{58} \times \theta^{C1}OH$ |
| 59 | $^{M}OH–{}^{C1}OH + H^+ + e^- \rightarrow {}^{M}OH + H_2O$ | $f_{59} \times \theta^{M}OH–{}^{C1}OH - b_{59} \times \theta^{M}OH$ |
| 60 | $^{C1}O + H^+ + e^- \rightarrow {}^{C1}OH$ | $f_{60} \times \theta^{C1}O - b_{60} \times \theta^{C1}OH$ |
| 61 | $^{C1}OH + H^+ + e^- \rightarrow M + H_2O$ | $f_{61} \times \theta^{C1}OH - b_{61} \times \theta^{M}$ |
| 62 | $^{M}O–{}^{C1}OH + H^+ + e^- \rightarrow {}^{M}O + H_2O$ | $f_{62} \times \theta^{M}O–{}^{C1}OH - b_{62} \times \theta^{M}O$ |
| 63 | $^{M}OOH–{}^{C1}OH + H^+ + e^- \rightarrow {}^{M}OOH + H_2O$ | $f_{63} \times \theta^{M}OOH–{}^{C1}OH - b_{63} \times \theta^{M}OOH$ |
| 64 | $^{M}O_2 + {}^\bullet OH \rightarrow {}^{M}O_2–{}^{C1}OH$ | $f_{64} \times \theta^{M}O_2 \times x({}^\bullet OH) - b_{64} \times \theta^{M}O_2–{}^{C1}OH$ |
| 65 | $^{C1}OH + {}^\bullet OH \rightarrow {}^{M}OH–{}^{C1}OH$ | $f_{65} \times \theta^{C1}OH \times x({}^\bullet OH) - b_{65} \times \theta^{M}OH–{}^{C1}OH$ |
| 66 | $^{M}H + {}^\bullet OH \rightarrow M + H_2O$ | $f_{66} \times \theta^{M}H \times x({}^\bullet OH) - b_{66} \times \theta^{M}$ |
| 67 | $^{M}O_2–{}^{C1}OH + H^+ + e^- \rightarrow {}^{M}O_2 + H_2O$ | $f_{67} \times \theta^{M}O_2–{}^{C1}OH - b_{67} \times \theta^{M}O_2$ |
| 68 | $^{M}O–{}^{M}OH + H^+ + e^- \rightarrow {}^{M}O + H_2O$ | $f_{68} \times \theta^{M}O–{}^{M}OH - b_{68} \times \theta^{M}O$ |
| 69 | $^M2OH + H^+ + e^- \rightarrow {}^{M}OH + H_2O$ | $f_{69} \times \theta^{M}2OH - b_{69} \times \theta^{M}OH$ |
| 70 | $^M3OH + H^+ + e^- \rightarrow {}^M2OH + H_2O$ | $f_{70} \times \theta^{M}3OH - b_{70} \times \theta^{M}2OH$ |
| | Interconversion between reactive oxygen species[11] | |
| 71 | $H_2O_2\,(aq) + {}^\bullet OH \rightarrow H_2O\,(l) + {}^\bullet OOH$ | $4.2 \times 10^7 \times x(H_2O_2) \times x({}^\bullet OH)$ |
| 72 | $H_2O_2\,(aq) + {}^\bullet OOH \rightarrow H_2O\,(l) + {}^\bullet OH + O_2\,(g)$ | $0.5 \times x(H_2O_2) \times x({}^\bullet OOH)$ |
| 73 | ${}^\bullet OH + {}^\bullet OH \rightarrow H_2O_2\,(aq)$ | $6.0 \times 10^9 \times x({}^\bullet OH) \times x({}^\bullet OH)$ |
| 74 | ${}^\bullet OOH + {}^\bullet OOH \rightarrow H_2O_2\,(aq) + O_2\,(g)$ | $9.8 \times 10^5 \times x({}^\bullet OOH) \times x({}^\bullet OOH)$ |
| 75 | ${}^\bullet OOH + {}^\bullet OH \rightarrow H_2O\,(l) + O_2\,(g)$ | $1.0 \times 10^{10} \times x({}^\bullet OOH) \times x({}^\bullet OH)$ |

$^M$, $^{C1}$, $^{N1}$, and $^{N2}$ represent the active site of MPc shown in **Figure 1b** of main text; *g*, *l*, and *aq* refer to the gas phase, liquid, and aqueous phase.

The time-dependent concentrations of species are governed by a set of coupled, non-linear ordinary differential equations (ODEs), as presented in eqs S1-S40. Each ODE describes the net rate for the corresponding species.

$$\frac{dx(O_2, dl)}{dt} = r_1 - r_2 - r_{55} \tag{S1}$$



$$\frac{d\theta^M}{dt} = -r_2 + r_6 + r_8 - r_{17} + r_{18} - r_{19} + r_{20} - r_{24} - r_{25} - r_{54} + r_{61} + r_{66} \tag{S2}$$

$$\frac{d\theta^{M}O_2}{dt} = r_2 - r_3 - r_9 + r_{10} - r_{64} + r_{67} \tag{S3}$$

$$\frac{d\theta^{M}OOH}{dt} = r_3 - r_4 - r_7 - r_{11} + r_{12} - r_{22} - r_{23} + r_{24} - r_{33} + r_{63} \tag{S4}$$

$$\frac{d\theta^{M}O}{dt} = r_4 - r_5 - r_{13} + r_{14} + r_{22} - r_{30} - r_{31} - r_{32} + r_{62} + r_{68} \tag{S5}$$

$$\frac{d\theta^{M}OH}{dt} = r_5 - r_6 - r_{15} + r_{16} + r_{21} + r_{25} - r_{26} - r_{27} - r_{28} + r_{59} + r_{69} \tag{S6}$$

$$\frac{d\theta^{M}H_2O_2}{dt} = r_7 - r_8 - r_{21} \tag{S7}$$

$$\frac{d\theta^{M}O_2-N^2H}{dt} = r_9 - r_{10} \tag{S8}$$

$$\frac{d\theta^{M}OOH-N^2H}{dt} = r_{11} - r_{12} \tag{S9}$$

$$\frac{d\theta^{M}O-N^2H}{dt} = r_{13} - r_{14} \tag{S10}$$

$$\frac{d\theta^{M}OH-N^2H}{dt} = r_{15} - r_{16} \tag{S11}$$

$$\frac{d\theta^{M}H}{dt} = r_{17} - r_{18} - r_{66} \tag{S12}$$

$$\frac{d\theta^{N^2H}}{dt} = r_{19} - r_{20} \tag{S13}$$

$$\frac{d\theta^{M}OH-C^1OOH}{dt} = r_{26} + r_{46} - r_{40} \tag{S14}$$

$$\frac{d\theta^{M}OH-C^1OH}{dt} = r_{27} + r_{45} - r_{36} - r_{58} - r_{59} + r_{65} \tag{S15}$$

$$\frac{d\theta^{M}O-C^1OOH}{dt} = r_{30} - r_{46} \tag{S16}$$

$$\frac{d\theta^{M}O-C^1OH}{dt} = r_{23} + r_{31} - r_{42} - r_{45} + r_{57} - r_{62} \tag{S17}$$

$$\frac{d\theta^{M}OH-C^1O}{dt} = r_{42} - r_{43} \tag{S18}$$

$$\frac{d\theta^{M}O-C^1O}{dt} = r_{39} + r_{40} - r_{41} \tag{S19}$$

$$\frac{d\theta^{M}OOH-C^1OH}{dt} = r_{33} - r_{34} - r_{39} + r_{56} - r_{57} - r_{63} \tag{S20}$$

$$\frac{d\theta^{M}H_2O_2-C^1O}{dt} = r_{34} - r_{35} \tag{S21}$$

$$\frac{d\theta^{M}H_2O-C^1O}{dt} = r_{36} - r_{37} \tag{S22}$$

$$\frac{d\theta^{C^1O}}{dt} = r_{35} + r_{37} - r_{38} - r_{60} \tag{S23}$$

$$\frac{d\theta^{C^1OH}}{dt} = r_{54} - r_{55} + r_{58} + r_{60} - r_{61} - r_{65} \tag{S24}$$

$$\frac{d\theta^{M}O_2-C^1OH}{dt} = r_{55} - r_{56} + r_{64} - r_{67} \tag{S25}$$

$$\frac{d\theta^{M}OH-O_{inter}}{dt} = r_{43} - r_{44} \tag{S26}$$

$$\frac{d\theta^{O_{inter}}}{dt} = r_{38} + r_{44} \tag{S27}$$

$$\frac{d\theta^{M}O-O_{outer}}{dt} = r_{41} \tag{S28}$$



$$\frac{d\theta^{M_O-M_{OH}}}{dt} = r_{32} - r_{47} - r_{68} \tag{S29}$$

$$\frac{d\theta^{M_{2OH}}}{dt} = r_{28} + r_{47} - r_{29} - r_{48} - r_{69} + r_{70} \tag{S30}$$

$$\frac{d\theta^{M_{2OH}-N^1H}}{dt} = r_{48} - r_{49} \tag{S31}$$

$$\frac{d\theta^{M_{2OH}-N^1H-N^3H}}{dt} = r_{49} - r_{50} \tag{S32}$$

$$\frac{d\theta^{M_{3OH}}}{dt} = r_{29} - r_{51} - r_{70} \tag{S33}$$

$$\frac{d\theta^{M_{3OH}-N^1H}}{dt} = r_{51} - r_{52} \tag{S34}$$

$$\frac{d\theta^{M_{3OH}-N^1H-N^3H}}{dt} = r_{52} - r_{53} \tag{S35}$$

$$\frac{d\theta^{Pc+M(OH)_2}}{dt} = r_{50} \tag{S36}$$

$$\frac{d\theta^{Pc+M(OH)_3}}{dt} = r_{53} \tag{S37}$$

$$\frac{dx(H_2O_2)}{dt} = r_8 - r_{10} - r_{12} - r_{14} - r_{16} - r_{18} - r_{20} + r_{35} - r_{71} - r_{72} + r_{73} + r_{74} \tag{S38}$$

$$\frac{dx(\cdot OOH)}{dt} = -r_{24} - r_{26} - r_{30} + r_{71} - r_{72} - 2r_{74} - r_{75} \tag{S39}$$

$$\frac{dx(\cdot OH)}{dt} = r_{10} + r_{12} + r_{14} + r_{16} + r_{18} + r_{20} + r_{21} + r_{22} - r_{25} - r_{27} - r_{28} - r_{29} - r_{31} - r_{32} - r_{33} - r_{54} - r_{64} - r_{65} - r_{66} - r_{71} + r_{72} - 2r_{73} - r_{75} \tag{S40}$$

For electrochemical steps, the forward and backward rate constant equations are shown in eqs S41-S42, respectively. Notably, the proton transfer from water to the reaction center in the electrochemical steps shares the same barrier value (0.26 eV) during ORR[12].

$$f_i = A_i e^{-\frac{E_{a,i} + (\Delta G_{i,0} + eU + 0.059 * pH * e)/2}{k_B T}} \tag{S41}$$

$$b_i = A_i e^{-\frac{E_{a,i} - (\Delta G_{i,0} + eU + 0.059 * pH * e)/2}{k_B T}} \tag{S42}$$

For chemical steps, the frequency factor is associated with vibrations of the bonds being broken or formed[13, 14], and generally ranges from $10^{12}$ to $10^{14}$ s$^{-1}$. The forward and backward rate constant equations are shown in eqs S43-S44, respectively.

$$f_i = \frac{k_B \times T}{h} e^{-\frac{E_{a,i}}{k_B T}} \tag{S43}$$

$$b_i = \frac{k_B \times T}{h} e^{-\frac{E_{a,i} - \Delta G_{i,0}}{k_B T}} \tag{S44}$$



## 3. Free energy diagrams for the 4e⁻ and 2e⁻ ORR processes on MPcs

**Figure S1** shows the free energy diagrams for the 4e⁻ and 2e⁻ processes on six MPcs. The limiting potential ($U_L$) for the 4e⁻ process decreases in the order of RhPc (0.69 V), IrPc (0.53 V), FePc (0.50 V), RuPc (0.43 V), MnPc (0.37 V), and CrPc (0.31 V). The $U_L$ for the 2e⁻ process decreases in the order of FePc (0.50 V), RhPc (0.37 V), MnPc (0.37 V), CrPc (-0.06 V), IrPc (-0.07 V), and RuPc (-0.15 V).

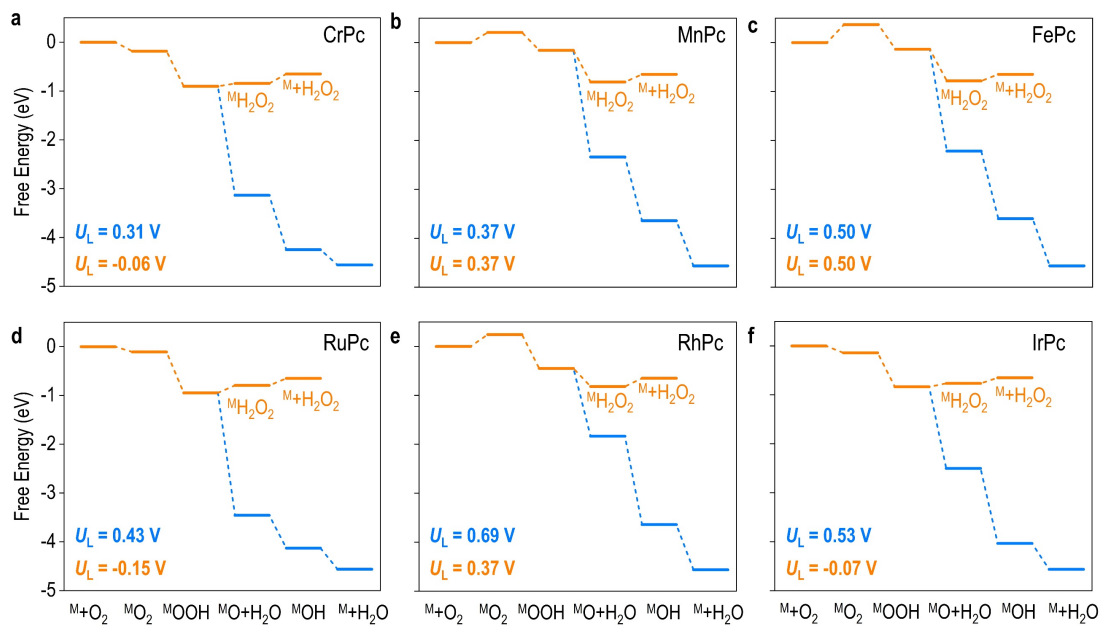

**Figure S1.** Free energy diagrams of ORR on MPcs for the 4e⁻ (blue) and 2e⁻ (orange) transfer processes at $T$ = 300 K, $P_{O2}$ = 1.0 atm, $P_{H2}$ = 1.0 atm, pH = 0, and $U$ = 0 V *vs*. SHE.



# 4. Microkinetic model with only the 4e⁻ and 2e⁻ ORR processes

**Table S3.** Kinetic and thermodynamic parameters for chemical steps and electrochemical steps.

| State $i$ | $A_i$ (s$^{-1}$)† | $E_{a,i}$ (eV)† | $\Delta G_i$ (eV) |
|---|---|---|---|
| -1: $O_2(aq)$ | $8 \times 10^5$ | 0 | 0 |
| 0: $O_2(dl)$ | $1 \times 10^8$ | 0 | $\Delta G_0 - \Delta G(O_2, \text{sol}, 300\text{K})$ |
| 1: $^M O_2$ | $1 \times 10^9$ | 0.26 | $\Delta G_1$ |
| 2: $^M OOH$ | $1 \times 10^9$ | 0.26 | $\Delta G_2$ |
| 3: $^M O$ | $1 \times 10^9$ | 0.26 | $\Delta G_3$ |
| 4: $^M OH$ | $1 \times 10^9$ | 0.26 | $\Delta G_4$ |
| 5: $^M H_2O_2$ | $1 \times 10^9$ | 0.26 | $\Delta G_5$ |
| 6: $H_2O_2(aq)$ | $1 \times 10^8$ | 0 | $\Delta G_6$ |
| 7: $^M$ | – | – | – |

*aq*: aqueous phase

*dl*: $O_2$ in electrical double layer

$\Delta G(O_2, \text{sol}, 300 \text{ K})$: 0.278 eV. The equilibrium mole fraction of $O_2$ in water, under 1 atm $O_2$ gas, was calculated to be $2.17 \times 10^{-5}$.

†: data from reference[15].

The free energy changes for reaction steps 0-6 are calculated with eqs S45-S51, respectively.

$$\Delta G_0 = G(*O_2) - G(*) - G(O_2, T, P) \tag{S45}$$

$$\Delta G_1 = G(*OOH) - G(*O_2) - 0.5G(H_2, T, P) + 2.303 \, k_B T \times \text{pH} - eU \tag{S46}$$

$$\Delta G_2 = G(*O) - G(*OOH) + G(H_2O) - 0.5G(H_2, T, P) + 2.303 \, k_B T \times \text{pH} - eU \tag{S47}$$

$$\Delta G_3 = G(*OH) - G(*O) - 0.5G(H_2, T, P) + 2.303 \, k_B T \times \text{pH} - eU \tag{S48}$$

$$\Delta G_4 = G(*) - G(*OH) + G(H_2O) - 0.5G(H_2, T, P) + 2.303 \, k_B T \times \text{pH} - eU \tag{S49}$$

$$\Delta G_5 = G(*H_2O_2) - G(*OOH) - 0.5G(H_2, T, P) + 2.303 \, k_B T \times \text{pH} - eU \tag{S50}$$

$$\Delta G_6 = G(*) + G(H_2O_2, T) - G(*H_2O_2) \tag{S51}$$



**Figure S2** illustrates the Pourbaix diagrams for the six MPcs with consideration of only the 4-electron and 2-electron processes. Across the examined range of pH (0 – 14) and potential (–0.5 to 0.4 V *vs.* SHE), bare surface ($^M$), $^M$OH, and $^M$O are the major surface states for CrPc and RuPc; $^M$H$_2$O$_2$, $^M$, and $^M$OH are the major surfaces states for MnPc, FePc and RhPc; and $^M$ and $^M$OH are the major surfaces states for IrPc.

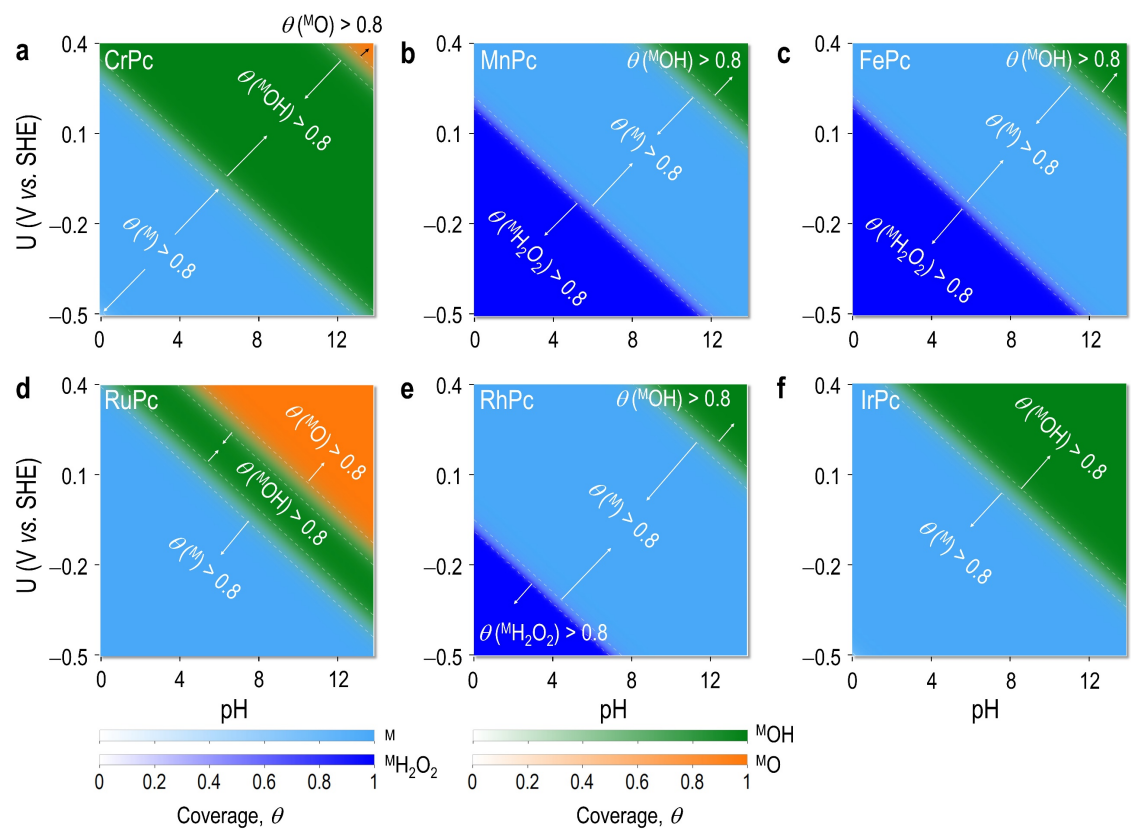

**Figure S2.** Pourbaix diagrams for MPcs obtained from steady-state microkinetic model with consideration of only the 4-electron and 2-electron ORR processes. The simulation conditions are $T$ = 300 K, $P_{O2}$ = 1.0 atm, and $P_{H2}$ = 1.0 atm.
S11

# 5. Generation and conversion of ROS in solution and on MPcs

**Table S4.** Generation and conversion of ROS on MPcs at $T$ = 300 K, $P_{O2}$ = 1 atm, $P_{H2}$ = 1 atm, pH = 0, and $U$ = 0 V *vs.* SHE. Types I, II, and III correspond to reactions involving with zero, one, and two radicals, respectively.

| ROS | Type | Reaction Step | dG (eV) for CrPc, MnPc, FePc, RuPc, RhPc, and IrPc |
|---|---|---|---|
| $O_2(g)$ | I | $2H_2O_2(aq) \rightarrow 2H_2O(l) + O_2(g)$ | -3.26 |
| | II | •OOH + $H_2O_2(aq)$ → •OH + $H_2O(l)$ + $O_2(g)$ [16,17] | -1.57 |
| | III | •OOH + •OH → $H_2O(l)$ + $O_2(g)$ [16] | -3.39 |
| | III | 2•OOH → $H_2O_2(aq)$ + $O_2(g)$ [16,17] | -1.69 |
| $H_2O_2(l)$ | I | $^MOOH + H^+ + e^- \rightarrow {}^MH_2O_2$ | 0.06, -0.65, -0.65, 0.15, -0.37, 0.07 |
| | I | $^MH_2O_2 \rightarrow {}^M + H_2O_2(aq)$ | 0.20, 0.16, 0.14, 0.15, 0.17, 0.12 |
| | II | •O + $H_2O(l)$ → $H_2O_2(aq)$ | -1.18 |
| | III | 2•OH → $H_2O_2(aq)$ [16,17] | -1.82 |
| | III | 2•OOH → $O_2(g) + H_2O_2(aq)$ [16,17] | -1.69 |
| •OH | I | $H_2O_2(aq) + {}^{N2}H \rightarrow {}^M + H_2O(l) + $ •OH [18] | -1.25, -1.32, -1.53, -0.98, -0.87, -0.74 |
| | I | $H_2O_2(aq) + {}^MH \rightarrow {}^M + H_2O(l) + $ •OH | -1.85, -2.33, -2.23, -1.28, -0.79, -0.28 |
| | I | $H_2O_2(aq) + {}^MO–{}^{N2}H \rightarrow {}^MO_2 + H_2O(l) + $ •OH | -1.23, -1.20, -1.28, -0.65, -0.85, -1.42 |
| | I | $H_2O_2(aq) + {}^MOOH–{}^{N2}H \rightarrow {}^MOOH + H_2O(l) + $ •OH [18] | -1.24, -1.12, -1.08, -0.96, -1.38, -1.43 |
| | I | $H_2O_2(aq) + {}^MO–{}^{N2}H \rightarrow {}^MO + H_2O(l) + $ •OH | -1.29, -1.11, -1.36, -1.34, -1.28, -1.37 |
| | I | $H_2O_2(aq) + {}^MOH–{}^{N2}H \rightarrow {}^MOH + H_2O(l) + $ •OH | -1.28, -1.25, -1.26, -0.95, -1.41, -1.47 |
| | I | $^MH_2O_2 \rightarrow {}^MOH + $ •OH | -0.54, 0.03, 0.05, -0.47, 0.05, -0.40 |
| | I | $^MOOH \rightarrow {}^MO + $ •OH | 0.63, 0.69, 0.79, 0.36, 1.48, 1.20 |
| | I | $H_2O_2(aq) \rightarrow 2$•OH [19] | 1.82 |
| | II | •OOH + $H_2O_2(aq)$ → •OH + $H_2O(l) + O_2(g)$ [16,17] | -1.57 |
| | II | •O + $H_2O(l)$ → 2•OH | 0.64 |
| | II | •OOH → •O + •OH | 2.88 |
| •OOH | I | $^M + H_2O_2(aq) \rightarrow {}^{N2}H + $ •OOH | 1.38, 1.44, 1.65, 1.10, 1.00, 0.87 |
| | I | $^M + H_2O_2(aq) \rightarrow {}^MH + $ •OOH | 1.97, 2.46, 3.57, 1.40, 0.91, 0.41 |
| | I | $H_2O_2(aq) \rightarrow $ •H + •OOH | 3.57 |
| | II | •H + $O_2(g)$ → •OOH [17] | -1.88 |
| | II | $H_2O_2(aq)$ + •OH → $H_2O(l)$ + •OOH | -1.70 |
| •O | I | $H_2O_2(aq) \rightarrow H_2O(l) + $ •O | 1.18 |
| | I | $^MO_2 \rightarrow {}^MO + $ •O | 2.15, 2.55, 2.51, 1.75, 3.02, 2.74 |
| | II | •OOH → •OH + •O | 2.88 |
| | III | 2•OH → $H_2O(l) + $ •O | -0.64 |
| •H | I | $^MH_2O_2 \rightarrow {}^MOOH + $ •H | 2.34, 3.05, 3.05, 2.25, 2.77, 2.04 |
| | I | $^MOOH \rightarrow {}^MO_2 + $ •H | 3.12, 2.77, 2.91, 3.24, 3.09, 3.10 |
| | I | $H_2O_2(aq) \rightarrow $ •OOH + •H | 3.57 |



# 6. Comparison between direct metal-leaching pathway and oxidation-induced demetallation processes

**Table S5.** Comparison between the direct metal-leaching pathway and stepwise oxidation-induced demetallation pathway for FePc.

| Reaction Pathway | ΔG (eV) | Initial State | Final State |
|---|---|---|---|
| FePc → FePc-with-displaced-Fe | 5.57 | 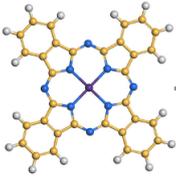 | 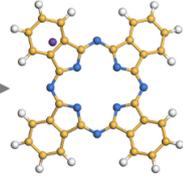 |
| state 23 → state 24 → state 25 → state 26 (Figure 2) | -1.19 (Barrier: 2.51 eV) | 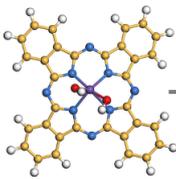 | 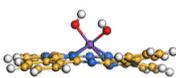 |



# 7. Comparison of adsorption ability at different sites of FePc

**Table S6.** Free energy changes of intermediates at different sites of FePc surface at $T$ = 300 K and $P_{H2}$ = 1atm. The energy unit is eV.

| Sites | $\Delta G(*OH)$ | $\Delta G(*OOH)$ | $\Delta G(*O)$ | $\Delta G(*H)$ |
|---|---|---|---|---|
| M | **0.96** | **4.43** | **2.35** | 1.19 |
| C1 | 1.53 | 4.89 | 2.83 | 0.77 |
| C2 | 3.11 | Desorb | 3.45 | 1.51 |
| C3 | 2.14 | 5.45 | 2.81 | 1.10 |
| C4 | 2.14 | 5.51 | 3.05 | 1.09 |
| N1 | change to M site | change to M site | 3.72 | 1.46 |
| N2 | 2.98 | Desorb | 2.72 | **0.48** |

The free energy changes on different sites are calculated with eqs S52-S55, respectively.

$$\Delta G(\,^*OH) = G(\,^*OH) - G(*) - G(H_2O, l) + 0.5\, G(H_2, g) \tag{S52}$$

$$\Delta G(\,^*OOH) = G(\,^*OOH) - G(*) - 2\, G(H_2O, l) + 1.5\, G(H_2, g) \tag{S53}$$

$$\Delta G(\,^*O) = G(\,^*O) - G(*) - G(H_2O, l) + G(H_2, g) \tag{S54}$$

$$\Delta G(\,^*H) = G(\,^*H) - G(*) - 0.5\, G(H_2, g) \tag{S55}$$

# 8. Radical attacks on the metal and C1 sites of MPcs

**Table S7.** Free energy changes for free radicals attacking the metal center and C1 sites on the MPcs surface at $T = 300$ K and $P_{H2} = 1$ atm. The energy unit is eV.

|  | CrPc | MnPc | FePc | RuPc | RhPc | IrPc |
|---|---|---|---|---|---|---|
| $\Delta G(^{M}OH)$ | -2.56 | -1.94 | -1.90 | -2.43 | -1.94 | -2.34 |
| $\Delta G(^{C1}OH)$ | -1.30 | -1.68 | -1.34 | -1.46 | -1.30 | -1.52 |
| $G(^{C1}OH)-G(^{M}OH)$ | 1.26 | 0.26 | 0.56 | 0.97 | 0.64 | 0.82 |
| $\Delta G(^{M}OOH)$ | -1.43 | -0.68 | -0.66 | -1.47 | -0.97 | -1.36 |
| $\Delta G(^{C1}OOH)$ | -0.16 | -0.21 | -0.19 | -0.29 | -0.08 | -0.30 |
| $G(^{C1}OOH)-G(^{M}OOH)$ | 1.27 | 0.47 | 0.47 | 1.18 | 0.89 | 1.06 |
| $\Delta G(^{M}O)$ | -3.67 | -2.87 | -2.75 | -3.99 | -2.37 | -3.04 |
| $\Delta G(^{C1}O)$ | -2.30 | -2.93 | -2.27 | -2.47 | -2.20 | -2.26 |
| $G(^{C1}O)-G(^{M}O)$ | 1.37 | -0.06 | 0.48 | 1.52 | 0.17 | 0.78 |

## 9. Comparison of protonation at different sites of MPcs

**Table S8.** Free energy change associated with protonation at different sites of MPcs under the conditions of $T$ = 300 K, $P_{H2}$ = 1.0 atm, pH = 0, and $U$ = 0 V *vs.* SHE. The energy unit is eV.

|  | $\Delta G$(Cr) | $\Delta G$(Mn) | $\Delta G$(Fe) | $\Delta G$(Ru) | $\Delta G$(Rh) | $\Delta G$(Ir) |
|---|---|---|---|---|---|---|
| $^M$H | 0.80 | 1.29 | 1.19 | 0.23 | -0.26 | -0.76 |
| $^{N2}$H | 0.21 | 0.27 | 0.48 | -0.07 | -0.17 | -0.30 |
| $^{C1}$H | 0.79 | 0.72 | 0.77 | 0.62 | 0.72 | 0.49 |
| $^{N1}$H | 1.32 | 1.49 | 1.46 | change to M site | 0.92 | change to M site |
| $^M$O$_2$–$^{N2}$H | 0.18 | 0.15 | 0.23 | -0.40 | -0.19 | 0.37 |
| $^M$OOH–$^{N2}$H | 0.20 | 0.08 | 0.04 | -0.09 | 0.34 | 0.38 |
| $^M$O–$^{N2}$H | 0.25 | 0.06 | 0.31 | 0.30 | 0.23 | 0.33 |
| $^M$OH–$^{N2}$H | 0.23 | 0.21 | 0.22 | -0.10 | 0.36 | 0.42 |

## 10. Variation of major byproduct distribution with pH and potential

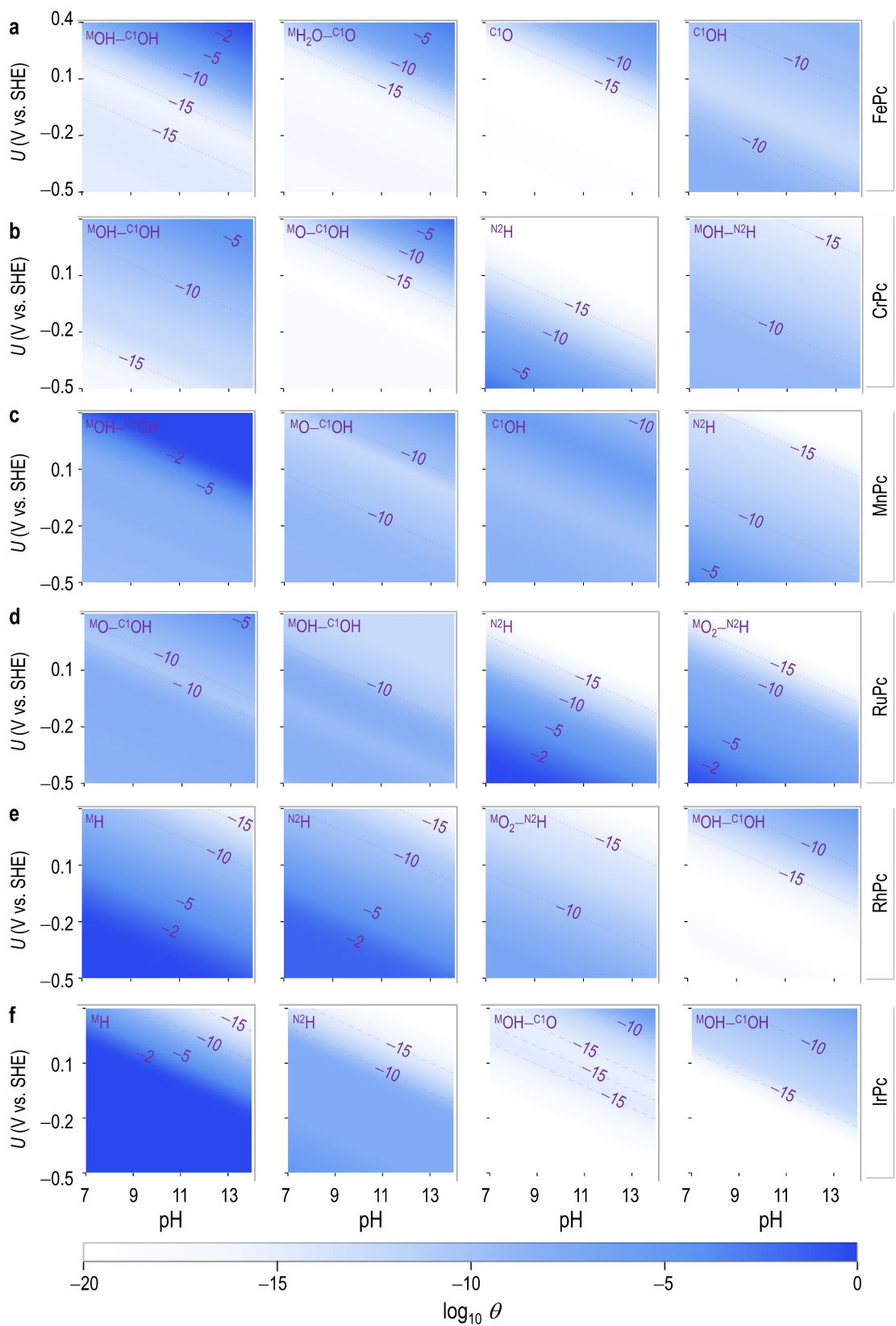

**Figure S3.** Pourbaix diagrams for the top four byproducts-I in each MPc after 3600 s.

S17

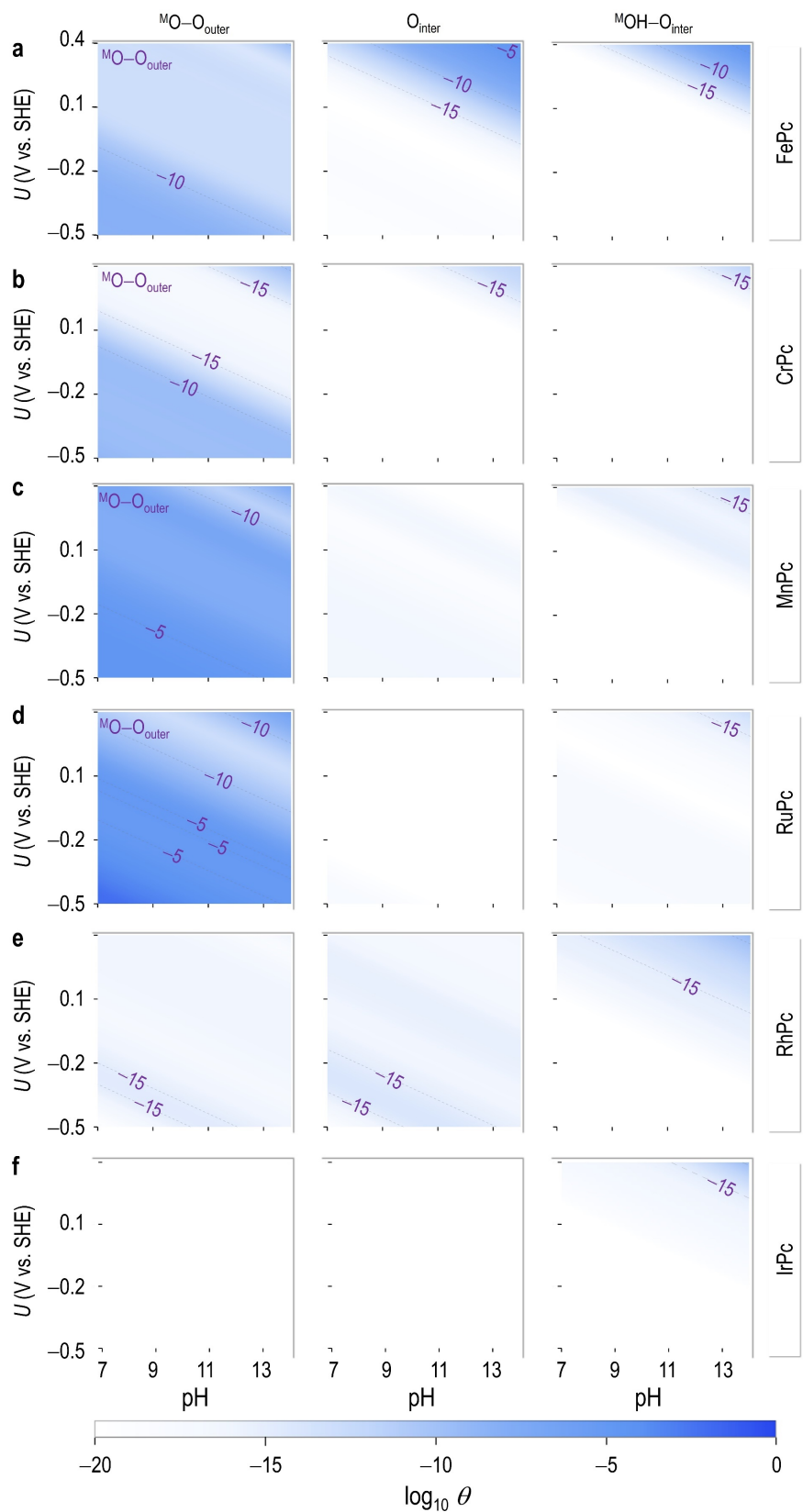

**Figure S4.** Pourbaix diagrams for the major byproducts-II in each MPc after 3600 s.

## 11. Reaction network of each MPc under typical operating conditions

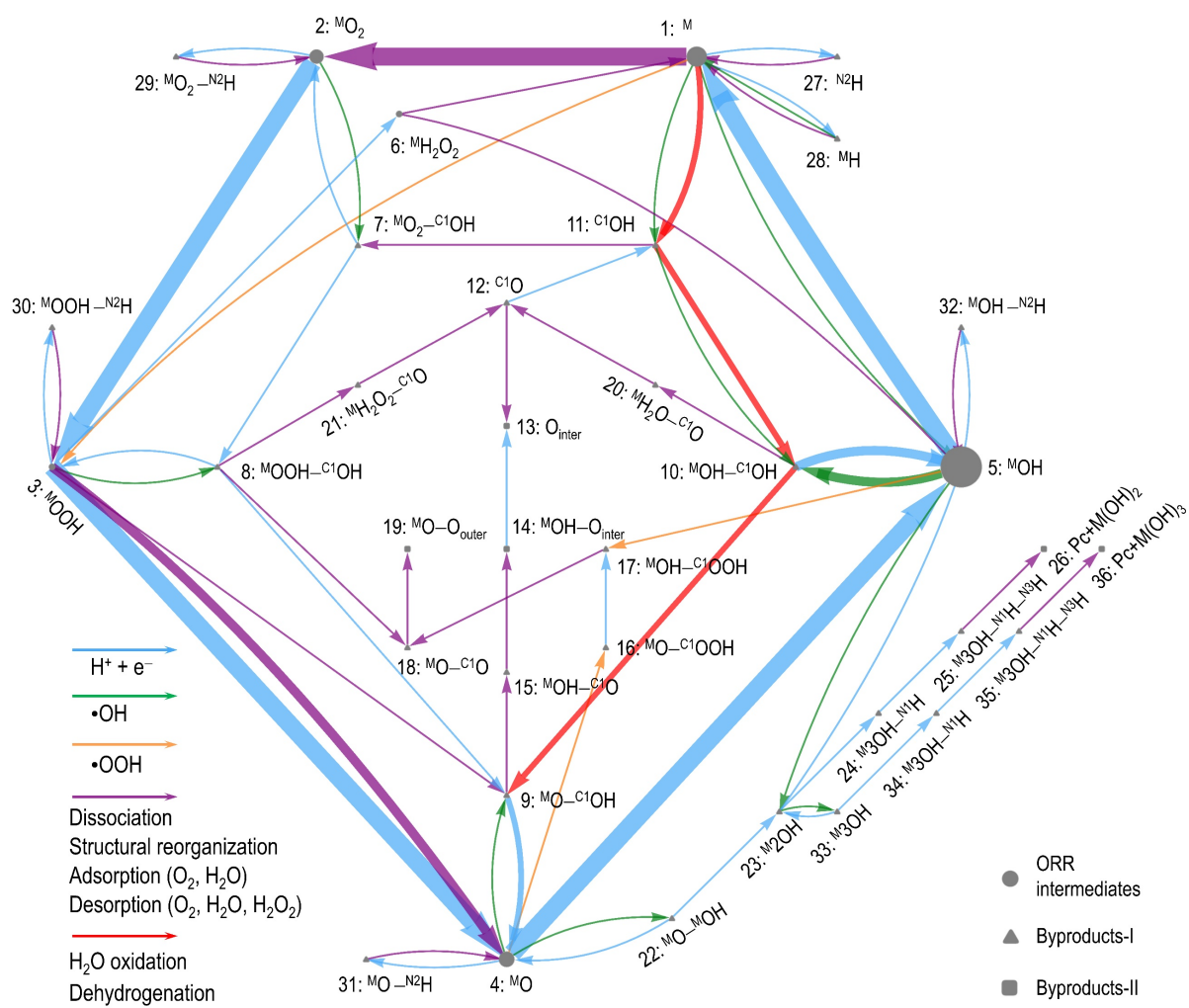

**Figure S5.** Reaction network for CrPc during ORR at pH 13 and −0.1 V *vs.* SHE (0.67 V *vs.* RHE) after 3600 s. Line thickness and node size represent the net reaction rate $r$ and surface coverage $\theta$, scaled by $2\log_{10}|r| + 25$ and $0.06\log_{10}\theta + 0.7$, respectively. For visualization clarity, $|r|$ and $\theta$ below $10^{-10}$ are assigned fixed minimum values: a line thickness of 2 and a fixed size of 0.1.

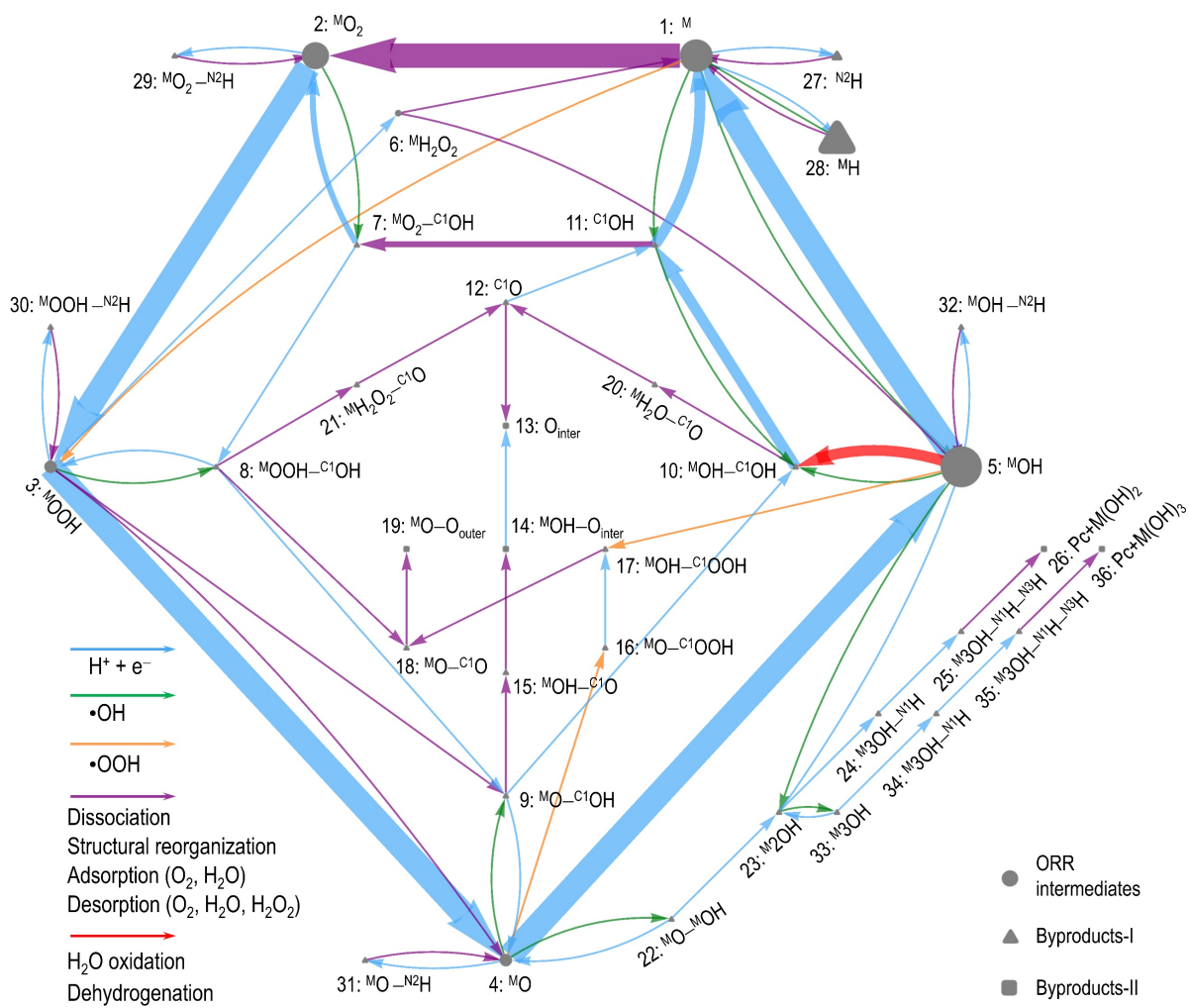

**Figure S6.** Reaction network for IrPc during ORR at pH 13 and −0.1 V *vs.* SHE (0.67 V *vs.* RHE) after 3600 s. Line thickness and node size represent the net reaction rate $r$ and surface coverage $\theta$, scaled by $2 \log_{10} |r| + 25$ and $0.06 \log_{10} \theta + 0.7$, respectively. For visualization clarity, $|r|$ and $\theta$ below $10^{-10}$ are assigned fixed minimum values: a line thickness of 2 and a fixed size of 0.1.

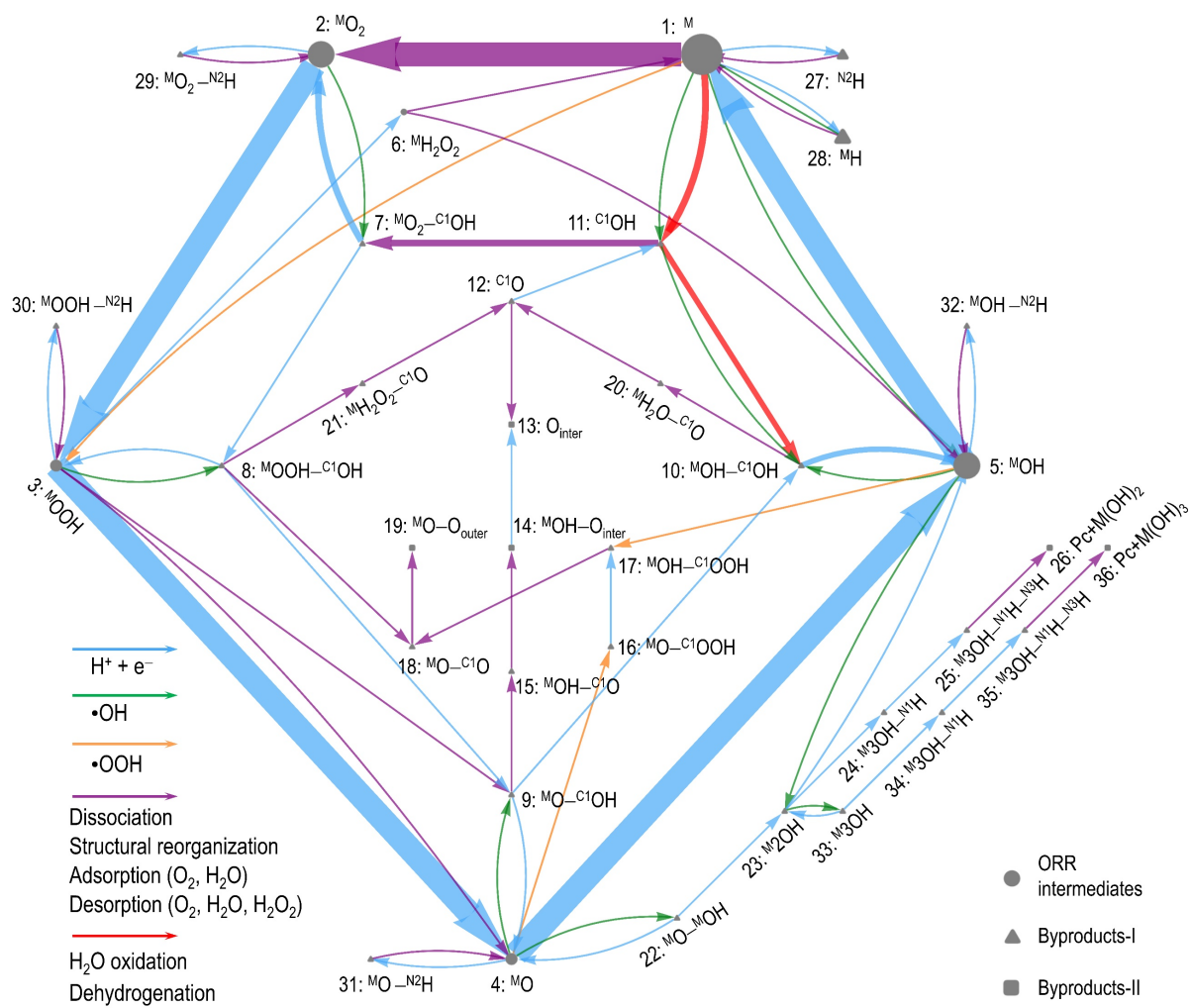

**Figure S7.** Reaction network for RhPc during ORR at pH 13 and −0.1 V *vs.* SHE (0.67 V *vs.* RHE) after 3600 s. Line thickness and node size represent the net reaction rate $r$ and surface coverage $\theta$, scaled by $2 \log_{10} |r| + 25$ and $0.06 \log_{10} \theta + 0.7$, respectively. For visualization clarity, $|r|$ and $\theta$ below $10^{-10}$ are assigned fixed minimum values: a line thickness of 2 and a fixed size of 0.1.

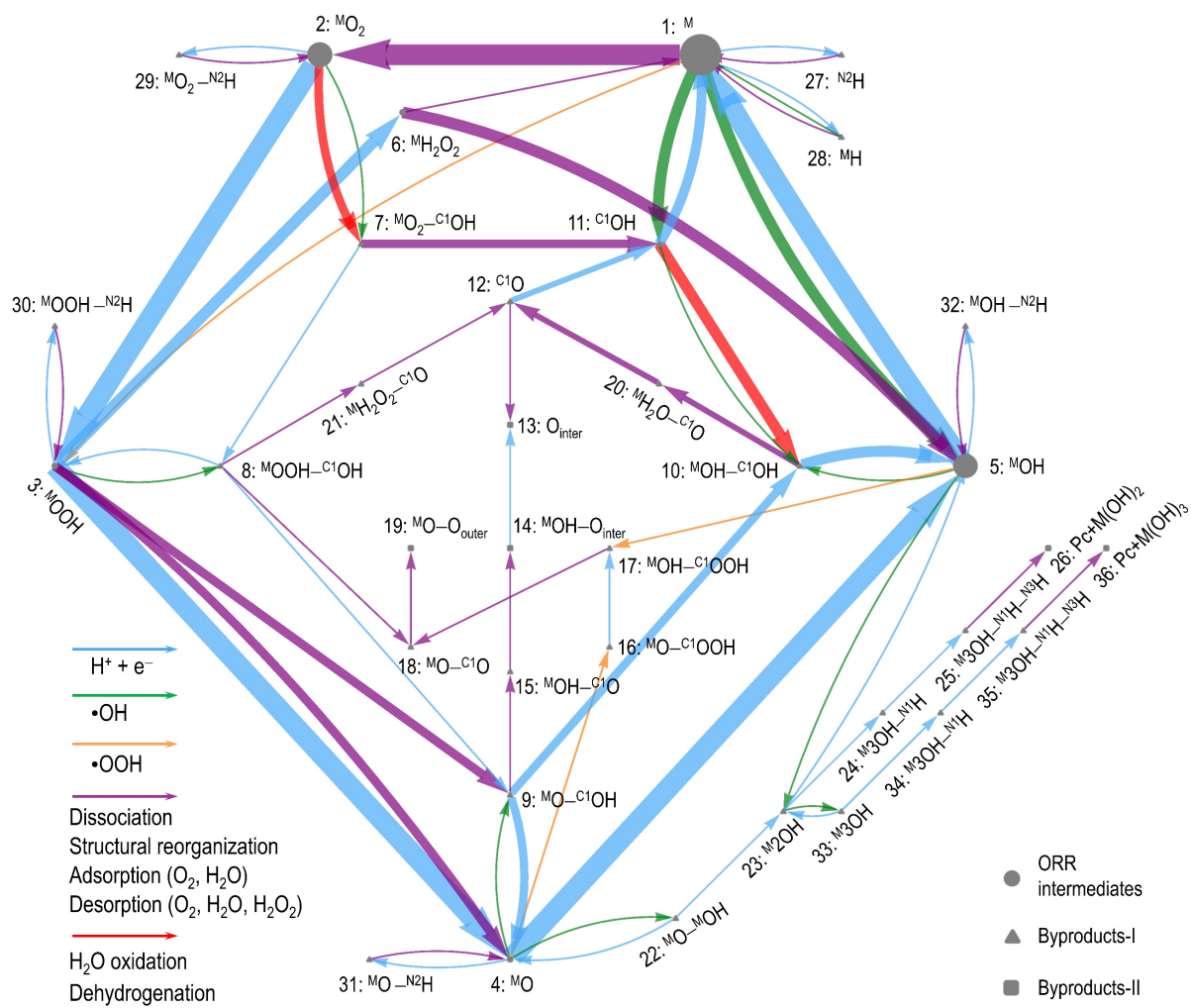

**Figure S8.** Reaction network for FePc during ORR at pH 13 and −0.1 V *vs.* SHE (0.67 V *vs.* RHE) after 3600 s. Line thickness and node size represent the net reaction rate $r$ and surface coverage $\theta$, scaled by $2\log_{10}|r| + 25$ and $0.06\log_{10}\theta + 0.7$, respectively. For visualization clarity, $|r|$ and $\theta$ below $10^{-10}$ are assigned fixed minimum values: a line thickness of 2 and a fixed size of 0.1.

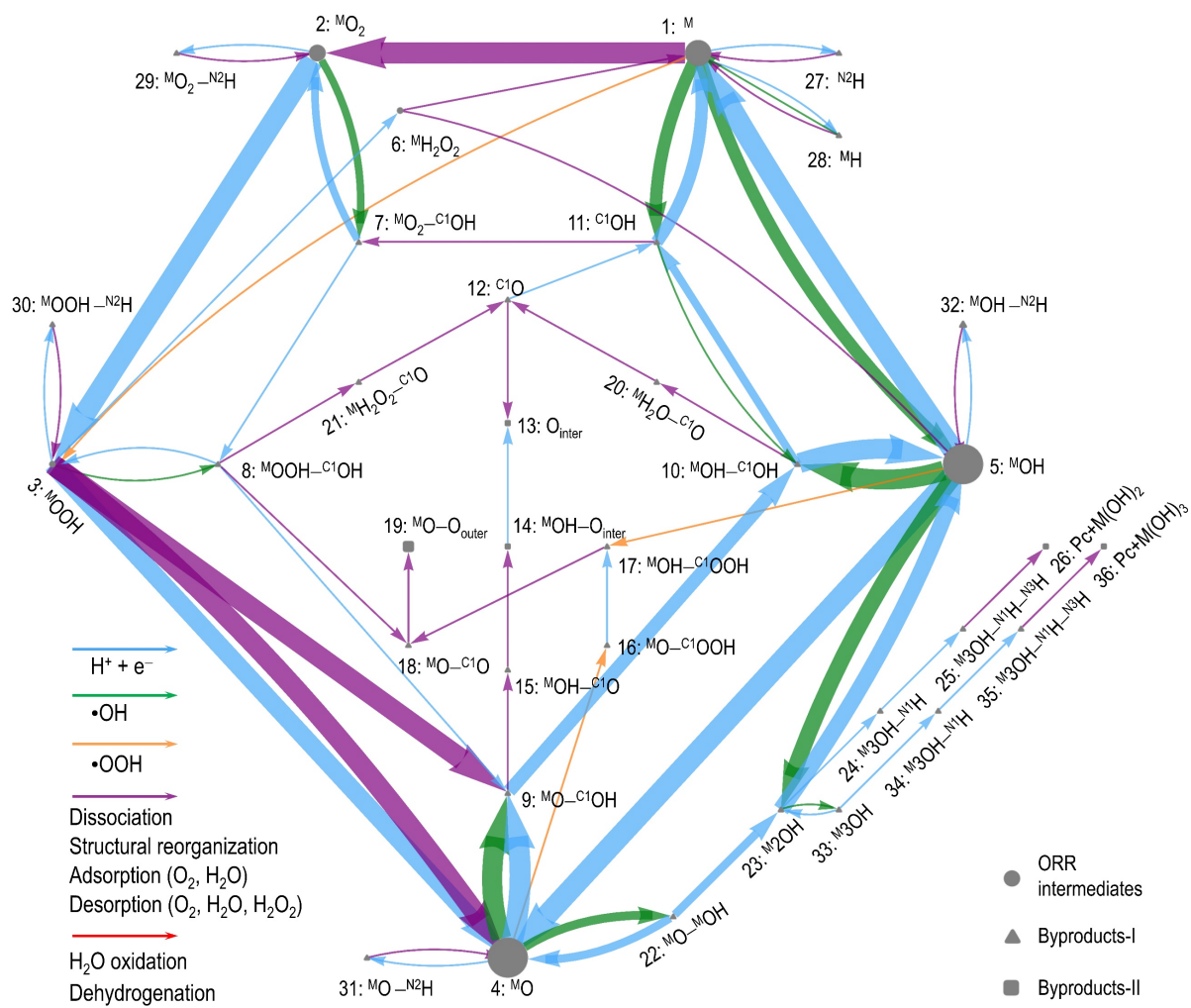

**Figure S9.** Reaction network for RuPc during ORR at pH 13 and −0.1 V *vs.* SHE (0.67 V *vs.* RHE) after 3600 s. Line thickness and node size represent the net reaction rate $r$ and surface coverage $\theta$, scaled by $2 \log_{10} |r| + 25$ and $0.06 \log_{10} \theta + 0.7$, respectively. For visualization clarity, $|r|$ and $\theta$ below $10^{-10}$ are assigned fixed minimum values: a line thickness of 2 and a fixed size of 0.1.

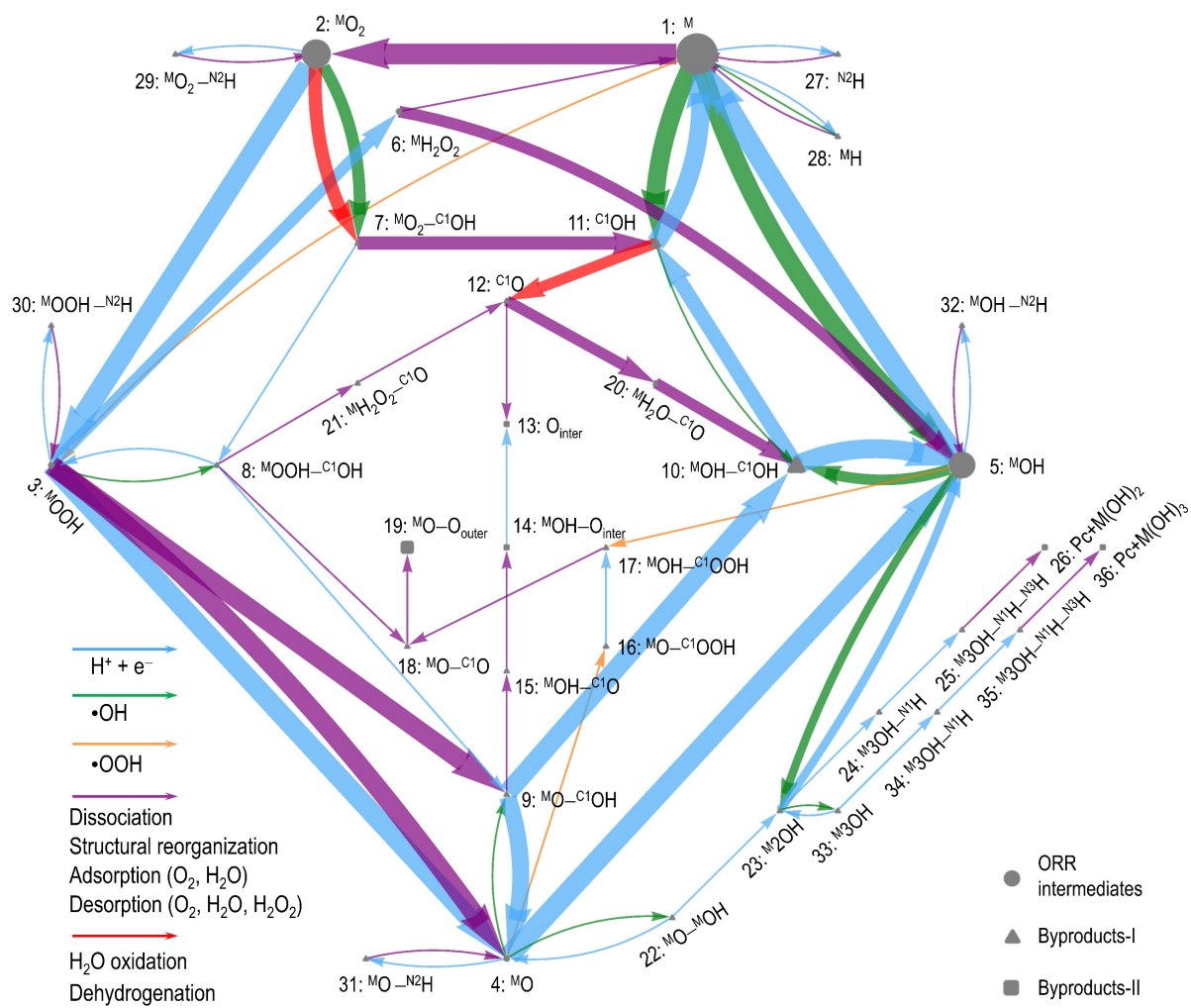

**Figure S10.** Reaction network for MnPc during ORR at pH 13 and −0.1 V *vs.* SHE (0.67 V *vs.* RHE) after 3600 s. Line thickness and node size represent the net reaction rate *r* and surface coverage $\theta$, scaled by $2 \log_{10} |r| + 25$ and $0.06 \log_{10} \theta + 0.7$, respectively. For visualization clarity, $|r|$ and $\theta$ below $10^{-10}$ are assigned fixed minimum values: a line thickness of 2 and a fixed size of 0.1.

**References**

(1) Yu, X.; Lai, S.; Xin, S.; Chen, S.; Zhang, X.; She, X.; Zhan, T.; Zhao, X.; Yang, D. Coupling of iron phthalocyanine at carbon defect site via π-π stacking for enhanced oxygen reduction reaction. *Appl. Catal. B: Environ.* **2021**, *280*, 119437. DOI: 10.1016/j.apcatb.2020.119437.

(2) Zhang, W.; Wang, L.; Zhang, L. H.; Chen, D.; Zhang, Y.; Yang, D.; Yan, N.; Yu, F. Creating hybrid coordination environment in Fe-based single atom catalyst for efficient oxygen reduction. *ChemSusChem* **2022**, *15* (12), e202200195. DOI: 10.1002/cssc.202200195.

(3) Zhang, W.; Meeus, E. J.; Wang, L.; Zhang, L. H.; Yang, S.; de Bruin, B.; Reek, J. N. H.; Yu, F. Boosting electrochemical oxygen reduction performance of iron phthalocyanine through axial coordination sphere interaction. *ChemSusChem* **2022**, *15* (3), e202102379. DOI: 10.1002/cssc.202102379.

(4) Liu, D.; Long, Y.-T. Superior catalytic activity of electrochemically reduced graphene oxide supported iron phthalocyanines toward oxygen reduction reaction. *ACS Appl. Mater. Inter.* **2015**, *7* (43), 24063-24068. DOI: 10.1021/acsami.5b07068.

(5) Zhang, Z. P.; Dou, M. L.; Ji, J.; Wang, F. Phthalocyanine tethered iron phthalocyanine on graphitized carbon black as superior electrocatalyst for oxygen reduction reaction. *Nano Energy* **2017**, *34*, 338-343. DOI: 10.1016/j.nanoen.2017.02.042.

(6) Mukherjee, M.; Samanta, M.; Banerjee, P.; Chattopadhyay, K. K.; Das, G. P. Endorsement of manganese phthalocyanine microstructures as electrocatalyst in ORR: Experimental and computational study. *Electrochim. Acta* **2019**, *296*, 528-534. DOI: 10.1016/j.electacta.2018.11.043.

(7) Li, J.; Chen, S.; Yang, N.; Deng, M.; Ibraheem, S.; Deng, J.; Li, J.; Li, L.; Wei, Z. Ultrahigh-loading zinc single-atom catalyst for highly efficient oxygen reduction in both acidic and alkaline media. *Angew. Chem. Int. Ed.* **2019**, *58* (21), 7035-7039. DOI: 10.1002/anie.201902109.

(8) Wen, X.; Yu, C.; Yan, B.; Zhang, X.; Liu, B.; Xie, H.; Kang Shen, P.; Qun Tian, Z. Morphological and microstructural engineering of Mn-N-C with strengthened Mn-N bond for efficient electrochemical oxygen reduction reaction. *Chem. Eng. J.* **2023**, *475*, 146135. DOI: 10.1016/j.cej.2023.146135.

(9) Gao, H.; Wang, Y.; Li, W.; Zhou, S.; Song, S.; Tian, X.; Yuan, Y.; Zhou, Y.; Zang, J. One-step carbonization of ZIF-8 in Mn-containing ambience to prepare Mn, N co-doped porous carbon as efficient oxygen reduction reaction electrocatalyst. *Int. J. Hydrogen Energy* **2021**, *46* (74), 36742-36752. DOI: 10.1016/j.ijhydene.2021.08.220.

(10) Wang, Y.; Zhang, X.; Xi, S.; Xiang, X.; Du, Y.; Chen, P.; Lyu, D.; Wang, S.; Tian, Z. Q.; Shen, P. K. Rational design and synthesis of hierarchical porous Mn–N–C nanoparticles with atomically dispersed MnN$_x$ moieties for highly efficient oxygen reduction reaction. *ACS Sustainable Chem. Eng.* **2020**, *8* (25), 9367-9376. DOI: 10.1021/acssuschemeng.0c01882.

(11) Yu, X.-Y.; Barker, J. R. Hydrogen peroxide photolysis in acidic aqueous solutions containing chloride ions. I. chemical mechanism. *J. Phys. Chem. A* **2003**, *107* (9), 1313-1324. DOI: 10.1021/jp0266648.

(12) Tripković, V.; Skúlason, E.; Siahrostami, S.; Nørskov, J. K.; Rossmeisl, J. The oxygen reduction reaction mechanism on Pt(111) from density functional theory calculations. *Electrochim. Acta* **2010**, *55* (27), 7975-7981. DOI: 10.1016/j.electacta.2010.02.056.

(13) He, Z. D.; Chen, Y. X.; Santos, E.; Schmickler, W. The Pre-exponential factor in electrochemistry. *Angew. Chem. Int. Ed.* **2018**, *57* (27), 7948-7956. DOI: 10.1002/anie.201800130.

(14) Skúlason, E.; Karlberg, G. S.; Rossmeisl, J.; Bligaard, T.; Greeley, J.; Jónsson, H.; Nørskov, J. K. Density

functional theory calculations for the hydrogen evolution reaction in an electrochemical double layer on the Pt(111) electrode. *Phys. Chem. Chem. Phys.* **2007**, *9* (25), 3241-3250. DOI: 10.1039/b700099e.

(15) Hansen, H. A.; Viswanathan, V.; Nørskov, J. K. Unifying kinetic and thermodynamic analysis of 2 e⁻ and 4 e⁻ reduction of oxygen on metal surfaces. *J. Phys. Chem. C* **2014**, *118* (13), 6706-6718. DOI: 10.1021/jp4100608.

(16) Pandis, S. N.; Seinfeld, J. H. Sensitivity analysis of a chemical mechanism for aqueous-phase atmospheric chemistry. *J. Geophys. Res.: Atmos.* **2012**, *94* (D1), 1105-1126. DOI: 10.1029/JD094iD01p01105.

(17) Danilczuk, M.; Coms, F. D.; Schlick, S. Visualizing chemical reactions and crossover processes in a fuel cell inserted in the ESR resonator: Detection by spin trapping of oxygen radicals, nafion-derived fragments, and hydrogen and deuterium atoms. *J. Phys. Chem. B* **2009**, *113* (23), 8031-8042. DOI: 10.1021/jp901597f.

(18) Zeng, H.; Zhang, G.; Ji, Q.; Liu, H.; Hua, X.; Xia, H.; Sillanpaa, M.; Qu, J. pH-independent production of hydroxyl radical from atomic H*-mediated electrocatalytic $H_2O_2$ reduction: A green Fenton process without byproducts. *Environ. Sci. Technol.* **2020**, *54* (22), 14725-14731. DOI: 10.1021/acs.est.0c04694.

(19) Schumb, W.; Satterfield, C. N.; Wentworth, R. L. *Hydrogen peroxide*; Reinhold Publishing Corporation, 1955. DOI: 10.1002/jps.3030450224.